\documentclass{article}
\usepackage{amssymb}
\usepackage{amsmath}
\usepackage{amsfonts}
\usepackage{sw20ssur}

\newtheorem{theorem}{Theorem}
\newtheorem{acknowledgement}[theorem]{Acknowledgement}

\input{tcilatex}
\begin{document}

\title{Constructing quantum games from non-factorizable joint probabilities}
\author{Azhar Iqbal$^{\text{a,b}}$ and Taksu Cheon$^{\text{a}}$ \\
$^{\text{a}}${\small Kochi University of Technology, Tosa Yamada, Kochi
782-8502, Japan.}\\
$^{\text{b}}${\small Centre for Advanced Mathematics and Physics,}\\
{\small National University of Sciences \& Technology,}\\
{\small campus of College of Electrical \& Mechanical Engineering,}\\
{\small Peshawar Road, Rawalpindi, Pakistan.}}
\maketitle
\tableofcontents

\begin{abstract}
A probabilistic framework is developed that gives a unifying perspective on
both the classical and the quantum games. We suggest exploiting peculiar
probabilities involved in Einstein-Podolsky-Rosen (EPR) experiments to
construct quantum games. In our framework a game attains classical
interpretation when joint probabilities are factorizable and a quantum game
corresponds when these probabilities cannot be factorized. We analyze how
non-factorizability changes Nash equilibria in two-player games while
considering the games of Prisoner's Dilemma, Stag Hunt, and Chicken. In this
framework we find that for the game of Prisoner's Dilemma even
non-factorizable EPR joint probabilities cannot be helpful to escape from
the classical outcome of the game. For a particular version of the Chicken
game, however, we find that the two non-factorizable sets of joint
probabilities, that maximally violates the Clauser-Holt-Shimony-Horne (CHSH)
sum of correlations, indeed result in new Nash equilibria.
\end{abstract}

\newpage

\section{Introduction}

Usual approach in the area of quantum games \cite%
{MeyerDavid,EWL,Multiplayer,Du,Cheon,Flitney,Shimamura,Iqbal,Ichikawa}
consists of analyzing a quantum system manoeuvred by participating agents,
recognized as players, who possess necessary means for their actions on
parts of the system. The quantum system evolves to its final state and
players' payoffs, or utilities, mathematically expressed as expectation
values of self-adjoint payoff operators, are generated from quantum
measurement \cite{Peres}. Thus the usual constructions of quantum games
involve the concepts of quantum state vectors, entangled states, quantum
measurement, expectation values, trace operation, and that of density
operators etc. This may seem normal because as being part of the research
field of quantum computation \cite{Nielsen} quantum games are expected to
exploit relevant tools from quantum mechanics. However, in our experience,
this noticeable reliance of the models of quantum games on the tools of
quantum mechanics also succeeds to keep many readers away from this
inter-disciplinary area of research. Ideally, they would like to see genuine
quantum games constructed from elementary probabilistic concepts, as it is
the case with many examples in game theory \cite{Rasmusen}. We find this
situation as an opportunity to present a probabilistic approach in which
quantum games are constructed without referring to the tools of quantum
mechanics.

While looking for the possibility of such an approach, it is encouraging to
find that the most unusual character of quantum mechanics can be expressed
in terms of probabilities \cite{Chung} only. For example, Bell inequalities 
\cite{Bell,Bell1,Bell2,Peres} can be written in terms of constraints on
joint probabilities relevant to pairs of certain random variables. As
probabilities are central to usual analyses in game theory, it seems natural
to use the peculiar probabilities, responsible for the violation of Bell
inequalities, to construct quantum games. We, therefore, suggest to
construct quantum games from the probabilities arising in the
Einstein-Podolsky-Rosen (EPR) experiments \cite{EPR,Bell,Bell2,Aspect et
al,Peres} performed to test the violation of Bell inequalities. The most
unusual character of the EPR probabilities, that they may not be
factorizable, motivates us, in this paper, to find how non-factorizability
can be used to construct quantum games. In other words, we search for the
role of non-factorizable probabilities in game-theoretic solution concepts
when EPR experiments provide the sets of non-factorizable probabilities.

This explicitly probabilistic approach towards quantum games is expected to
be of interest to the readers from such areas as economics \cite{Aumann} and
mathematical biology \cite{Hofbauer Sigmund}, where game theory finds
extensive applications and the tools of quantum mechanics are found to be
rather alien. Secondly, because of its exclusively probabilistic content,
this approach promises to provide a unified perspective for both the
classical and the quantum games.

The rest of the paper is organized as follows. Section \ref{FvBI} describes
the role of factorizability in deriving Bell inequality. Section \ref{tpgFPs}
discusses playing two-player games using factorizable probabilities and
presents two- and four-coin setups to play the well known games of
Prisoner's Dilemma (PD), Stag Hunt (SH), and Chicken \cite{Rasmusen}.
Section \ref{tpguEPRs} describes playing two-player games with EPR
experiments. Section \ref{tpgunfps} develops a framework in which
factorizable probabilities lead to the classical game whereas
non-factorizable probabilities result in the quantum game. Section \ref%
{Discussion} discusses the results and presents a view for further work.

\section{\label{FvBI}Factorizability and violation of Bell inequality}

Factorizability is known to be an interesting property of coupled systems
with separated parts, saying \cite{Fine} that for such systems the
probability for a simultaneous pair of outcomes can be expressed as the
product of the probability for each outcome separately. Mathematically, this
is expressed by writing joint probabilities as arithmetic product of their
respective marginals \cite{WinsbergFine}. It turns out to be the most
important technical ingredient in the proof of Bell inequality. In fact, J.
S. Bell used factorizability in his mathematical formulation of locality 
\cite{Peres}. Others \cite{Fine,Fine1,WinsbergFine} have recognized it, for
example, by such terms as \textquotedblleft conditional stochastic
dependence.\textquotedblright

We describe factorizability with reference to the standard EPR setup \cite%
{Aspect et al,Peres} used to derive Bell inequality. This setup consists of
two spatially-separated participants, known as Alice and Bob, who share
two-particle systems emitted by the same source. We specify the state of a
coupled system by $\lambda $. We denote Alice's parameter by $a$ that can be
set either at $S_{1}$ or at $S_{2}$ and denote Bob's parameter by $b$ that
can be set either at $S_{1}^{\prime }$ or at $S_{2}^{\prime }$.

In a run, Alice sets her apparatus either at $S_{1}$ or at $S_{2}$ and, in
either case, on receiving her particle she makes a measurement, the outcome
of which is $\pi _{A}$ that is either $+1$ or $-1$. In the same run, Bob
sets his apparatus either at $S_{1}^{^{\prime }}$ or at $S_{2}^{\prime }$
and, in either case, on receiving his particle he makes a measurement, the
outcome of which is $\pi _{B}$ that can be either $+1$ or $-1$. Alice and
Bob record the outcomes of their measurements for many runs as they receive
two-particle systems emitted from the same source.

We denote the probability that Alice obtains the outcome $\pi _{A}=+1$ or $%
-1 $ by $\Pr (\lambda ;\pi _{A},a)$ and, similarly, we denote the
probability that Bob obtains the outcome $\pi _{B}=+1$ or $-1$ by $\Pr
(\lambda ;\pi _{B};b)$. Also, we denote the probability that Alice and Bob
obtain the outcomes $\pi _{A}$ and $\pi _{B}$, respectively, by $\Pr
(\lambda ;\pi _{A},\pi _{B};a,b)$. These outcomes result from their choices
of the parameters $a$ and $b$, i.e. which one of the four pairs $%
(S_{1},S_{1}^{\prime }),$ $(S_{1},S_{2}^{\prime }),$ $(S_{2},S_{1}^{\prime
}),$ $(S_{2},S_{2}^{\prime })$ is realized in a run.

According to quantum theory, the outcomes $\pi _{A}$ and $\pi _{B}$ both are
completely random and Alice and Bob can only find the probabilities to
obtain $+1$ or $-1$ at the outcomes of their measurements. As it is
described in the Section \ref{tpguEPRs}, Alice's and Bob's parameters $a$
and $b$ decide these probabilities.

In many runs, Alice can choose between $S_{1}$ or $S_{2}$ with some
probability. Similarly, in many runs, Bob can choose between $S_{1}^{\prime
} $ or $S_{2}^{\prime }$ with some probability.

Assume that the source emits a total of $N$ two-particle systems. We denote
by $N(\pi _{A};a)$ the number of times Alice gets the outcome $\pi _{A}$
when she may set her parameter $a$ either at $S_{1}$ or at $S_{2}$.
Similarly, we denote by $N(\pi _{B};b)$ the number of times Bob gets the
outcome $\pi _{B}$ when he may set his parameter $b$ either at $%
S_{1}^{\prime }$ or at $S_{2}^{\prime }$. And, we denote by $N(\pi _{A},\pi
_{B};a,b)$ the number of times when Alice gets the outcome $\pi _{A}$ and
Bob gets the outcome $\pi _{B}$, wherever they may set their parameters $a$
and $b$, respectively. When $N$ is large, the ensemble probabilities are
defined as

\begin{equation}
\begin{array}{l}
\Pr (\pi _{A};a)=N(\pi _{A};a)/N,\text{ \ \ }\Pr (\pi _{A};b)=N(\pi
_{B};b)/N, \\ 
\Pr (\pi _{A},\pi _{B};a,b)=N(\pi _{A},\pi _{B};a,b)/N.%
\end{array}%
\end{equation}%
For many runs we consider an ensemble of states emitted from the source that
may not be the same. To allow mixture of states we denote the normalized
probability density, characterizing the ensemble of emissions, by $\rho
(\lambda )$. The ensemble probabilities are

\begin{equation}
\begin{array}{l}
\Pr (\pi _{A};a)=\int_{\Gamma }\rho (\lambda )\Pr (\lambda ;\pi
_{A};a)d\lambda , \\ 
\Pr (\pi _{B};b)=\int_{\Gamma }\rho (\lambda )\Pr (\lambda ;\pi
_{B};b)d\lambda , \\ 
\Pr (\pi _{A},\pi _{B};a,b)=\int_{\Gamma }\rho (\lambda )\Pr (\lambda ;\pi
_{A},\pi _{B};a,b)d\lambda ,%
\end{array}%
\end{equation}%
where ${\small \Gamma }$ is the space of states $\lambda $. Now,
factorizability states that

\begin{equation}
\Pr (\lambda ;\pi _{A},\pi _{B};a,b)=\Pr (\lambda ;\pi _{A};a)\Pr (\lambda
;\pi _{B};b).  \label{factorizability (mathematical)}
\end{equation}%
That is, for each $\lambda $ the joint probabilities are arithmetic product
of their respective marginals.

Factorizability being the most important technical ingredient in the proof
of Bell inequality can be seen as follows. Following inequalities must hold
when probabilities are sensible quantities

\begin{equation}
0\leq \Pr (\lambda ;\pi _{A};a)\leq 1,\text{ \ \ }0\leq \Pr (\lambda ;\pi
_{B};b)\leq 1.  \label{Factorizability (a)}
\end{equation}%
We now refer to a theorem stating that if six numbers $\omega _{1}$, $\omega
_{2}$, $\varpi _{1}$, $\varpi _{2}$, $\varsigma _{1}$, $\varsigma _{2}$ are
given such that

\begin{equation}
0\leq \omega _{1}\leq \varsigma _{1},\text{ \ \ }0\leq \omega _{2}\leq
\varsigma _{1},\text{ \ \ }0\leq \varpi _{1}\leq \varsigma _{2},\text{ \ \ }%
0\leq \varpi _{2}\leq \varsigma _{2},  \label{Factorizability (b)}
\end{equation}%
then the function $\Xi =\omega _{1}\varpi _{1}-\omega _{1}\varpi _{2}+\omega
_{2}\varpi _{1}+\omega _{2}\varpi _{2}-\varsigma _{2}\omega _{2}-\varsigma
_{1}\varpi _{1}$ is constrained by the inequalities

\begin{equation}
-\varsigma _{1}\varsigma _{2}\leq \Xi \leq 0.  \label{Factorizability (c)}
\end{equation}%
Proof of this theorem can, for example, be found in Ref. \cite{Clauser Horne}%
. Inequalities (\ref{Factorizability (a)}) together with the inequalities (%
\ref{Factorizability (c)}) give

\begin{equation}
\begin{array}{l}
-1\leq \Pr (\lambda ;\pi _{A};S_{1})\Pr (\lambda ;\pi _{B};S_{1}^{\prime
})-\Pr (\lambda ;\pi _{A};S_{1})\Pr (\lambda ;\pi _{B};S_{2}^{\prime })+ \\ 
\Pr (\lambda ;\pi _{A};S_{2})\Pr (\lambda ;\pi _{B};S_{1}^{\prime })+\Pr
(\lambda ;\pi _{A};S_{2})\Pr (\lambda ;\pi _{B};S_{2}^{\prime })- \\ 
\Pr (\lambda ;\pi _{B};S_{2})-\Pr (\lambda ;\pi _{B};S_{1}^{\prime })\leq 0,%
\end{array}
\label{Factorizability (d)}
\end{equation}%
for each $\lambda $. The central role of factorizability follows as when the
definition (\ref{factorizability (mathematical)}) of factorizability holds,
the multiplication of (\ref{Factorizability (d)}) by $\rho (\lambda )$ and
the subsequent integration over $\lambda $ gives

\begin{equation}
\begin{array}{l}
-1\leq \Pr (\pi _{A},\pi _{B};S_{1},S_{1}^{\prime })-\Pr (\pi _{A},\pi
_{B};S_{1},S_{2}^{\prime })+\Pr (\pi _{A},\pi _{B};S_{2},S_{1}^{\prime })+
\\ 
\Pr (\pi _{A},\pi _{B};S_{2},S_{2}^{\prime })-\Pr (\pi _{A};S_{2})-\Pr (\pi
_{B};S_{1}^{\prime })\leq 0,%
\end{array}
\label{CHSH inequality}
\end{equation}%
which is the Bell (or Clauser-Horne) inequality \cite{Clauser Horne}. Note
that the definition (\ref{factorizability (mathematical)}) is crucial in
order to obtain the inequality (\ref{CHSH inequality}).

\section{\label{tpgFPs}Two-player games using factorizable probabilities}

To bring non-factorizability into the realm of two-player games, we consider
a symmetric two-player, two-strategy, non-cooperative game \cite{Rasmusen}
represented by the matrices

\begin{equation}
\mathcal{A}=%
\begin{array}{c}
\text{Alice}%
\end{array}%
\begin{array}{c}
X_{1} \\ 
X_{2}%
\end{array}%
\overset{\overset{%
\begin{array}{c}
\text{Bob}%
\end{array}%
}{%
\begin{array}{cc}
X_{1}^{\prime } & X_{2}^{\prime }%
\end{array}%
}}{\left( 
\begin{array}{cc}
K & L \\ 
M & N%
\end{array}%
\right) },\text{ \ \ }\mathcal{B}=%
\begin{array}{c}
\text{Alice}%
\end{array}%
\begin{array}{c}
X_{1} \\ 
X_{2}%
\end{array}%
\overset{\overset{%
\begin{array}{c}
\text{Bob}%
\end{array}%
}{%
\begin{array}{cc}
X_{1}^{\prime } & X_{2}^{\prime }%
\end{array}%
}}{\left( 
\begin{array}{cc}
K & M \\ 
L & N%
\end{array}%
\right) ,}  \label{A and B matrices}
\end{equation}%
where all $K,L,M,N$ are real numbers. Players can go for one of the two
available strategies: $X_{1}$, $X_{2}$ for Alice and $X_{1}^{\prime }$, $%
X_{2}^{\prime }$ for Bob.

As factorizability is central to obtain Bell inequality, in this paper we
construct quantum games from non-factorizable probabilities that exploit EPR
setup. This rests on Fine's view \cite{Fine} that the violation of Bell
inequality in EPR experiments shows that quantum theory violates
factorizability. This view allows us to construct quantum games for which
factorizability always corresponds to the classical game.

We recognize key features of an EPR setup being that these relate to a
probabilistic system divided into two parts such that a) each observer has
access to one part of the system b) each observer can select between two
available choices c) observers cannot communicate between themselves d)
observers can make independent selections between the available choices e)
probabilities relevant to each part of the system are normalized\footnote{%
Its exact meaning will be described shortly.} and that f) probabilities are
sensible quantities.

It is worth mentioning here that the experimental testing of Bell inequality
involves four correlation experiments that correspond to combining $S_{1}$
with $S_{1}^{\prime }$, $S_{1}$ with $S_{2}^{\prime }$, $S_{2}$ with $%
S_{1}^{\prime }$, and $S_{2}$ with $S_{2}^{\prime }$, respectively. These
experiments are mutually exclusive in the sense that for any given
experiment Alice has to select between $S_{1}$ and $S_{2}$ and Bob has to
select between $S_{1}^{\prime }$ and $S_{2}^{\prime }$. That is, Alice (Bob)
cannot go for $S_{1}$ $(S_{1}^{\prime })$ and $S_{2}$ $(S_{2}^{\prime })$
simultaneously because the corresponding observables are incompatible, and
cannot be measured simultaneously. Whereas, in the above derivation of the
Bell inequality it is assumed that $S_{1},S_{1}^{\prime
},S_{2},S_{2}^{\prime }$ all have definite values which can be measured
simultaneously in pairs.

\subsection{\label{pgwc}Games with coins}

The above mentioned features are remindful of coins which, if distributed
between players, are found to have all the above mentioned properties. For
coins factorizability has a straightforward meaning in that the associated
probabilities remain factorizable. Hence, we develop an analysis of
two-player games with non-factorizable probabilities by first translating
playing of three well known games in terms of the games played when players
share coins. It turns out that a version of this translation provides the
right comparison with the probabilities involved in the EPR experiments and
opens the way to the next step i.e. to introduce non-factorizable
probabilities into the playing of two-player games.

\subsubsection{Two-coin setup}

We now consider pairs of coins and use it to play a two-player game (\ref{A
and B matrices}). For example, this game can be played when each player
receives a coin, head up, and `to flip' or to `not to flip' is a player's
strategy. Both coins are then passed to a referee who rewards the players
after observing the state of both coins.

Assume $S_{1}$ (to flip) and $S_{2}$ (not to flip) are Alice's strategies
and $S_{1}^{\prime }$ (to flip) and $S_{2}^{\prime }$ (not to flip) are
Bob's strategies. That is, with reference to the matrices (\ref{A and B
matrices}), we make the association $S_{1}\sim X_{1}$, $S_{2}\sim X_{2}$,
and $S_{1}^{\prime }\sim X_{1}^{\prime }$, $S_{1}^{\prime }\sim
X_{1}^{\prime }$. In two-coin setup, we assume that the strategies $S_{1}$
and $S_{1}^{\prime }$ represent Alice's and Bob's actions `to flip' the
coin, respectively; and, similarly, $S_{2}$ and $S_{2}^{\prime }$ represent
Alice's and Bob's actions `not to flip' the coin, respectively.

In repeated runs of the game players can play mixed strategies. Alice's
mixed strategy $x$ $\in \lbrack 0,1]$ is the probability to choose $S_{1}$
over $S_{2}$ and similarly Bob's mixed strategy $y$ $\in \lbrack 0,1]$ is
the probability to choose $S_{1}^{\prime }$ over $S_{2}^{\prime }$. Players'
payoffs are written as

\begin{equation}
\Pi _{A,B}(x,y)=%
\begin{pmatrix}
x \\ 
1-x%
\end{pmatrix}%
^{T}%
\begin{pmatrix}
(K,K) & (L,M) \\ 
(M,L) & (N,N)%
\end{pmatrix}%
\begin{pmatrix}
y \\ 
1-y%
\end{pmatrix}%
,  \label{2-Coin payoff relations}
\end{equation}%
where $T$ is for transpose and the subscripts $A$ and $B$ refer to Alice and
Bob, respectively. The first and the second entries in a parentheses are
Alice's and Bob's payoffs, respectively. Assume $(x^{\star },y^{\star })$ is
a Nash equilibrium (NE) \cite{Rasmusen}:

\begin{equation}
\Pi _{A}(x^{\star },y^{\star })-\Pi _{A}(x,y^{\star })\geqslant 0,\text{ \ \ 
}\Pi _{B}(x^{\star },y^{\star })-\Pi _{B}(x^{\star },y)\geqslant 0.
\label{NE factorable 1}
\end{equation}%
In rest of this paper we will use \textquotedblleft NE\textquotedblright\
when we refer to either a Nash equilibrium or to Nash equilibria, assuming
that the right meaning can be judged from the context. We identify this
arrangement to play a two-player game using two coins as the \emph{two-coin
setup}.

\subsubsection{Four-coin setup}

The game (\ref{A and B matrices}) can also be played using four coins
instead of two. It is arranged by assigning two coins to each player before
the game is played. In a run each player has to choose one coin. Two coins
out of four are, therefore, chosen by the players in each turn. These coins
are then passed to a referee who tosses them together and observes the
outcome. It is assumed that the players do not need to share fair coins.

We recall that in two-coin setup $S_{1}$ and $S_{1}^{\prime }$ are Alice's
and Bob's strategies, respectively, that represent players' actions `to
flip' the coin that a player receives in a turn. Instead of flipping or not
flipping, in four-coin setup a player's strategy is to choose one out of the
two coins that are made available to each player in a turn. The four-coin
setup is relevant as, in a run, choosing a coin out of the two corresponds
to choosing one of the two directions in which measurement is performed in
standard EPR experiment, the outcome of which is $+1$ or $-1$.

In repeated game, a player's strategy is defined by the selection she/he
makes over several runs of the game. For example, a player plays a pure
strategy when she/he goes for the same coin over all the runs and plays a
mixed strategy when she/he finds a probability to choose one coin out of the
two over many runs. Referee rewards the players according to their
strategies, the underlying statistics of four coins obtained from the
outcomes of many tosses each one of which follows every time the two players
choose two out of the total four coins, and the matrices (\ref{A and B
matrices}) representing the game being played.

We identify the arrangement using four coins to play a two-player game as
the \emph{four-coin setup}. Note that in four-coin setup the players'
rewards depend on the outcomes of repeated tosses even for pure strategies.
A large number of runs are, therefore, necessary whether a player plays a
pure-strategy or a mixed-strategy. Four-coin setup provides an inherently
probabilistic character to playing a two-player game and facilitates a
probabilistic analysis when we seek to play the game (\ref{A and B matrices}%
) using EPR experiments.

As the four-coin setup uses a different definition of a strategy relative to
the two-coin setup, we call $S_{1}$ and $S_{2}$ being Alice's coins and $%
S_{1}^{\prime }$ and $S_{2}^{\prime }$ being Bob's coins. When selecting a
coin is a player's strategy and we want to play the game given by the
matrices (\ref{A and B matrices}), it is reasonable to make the association $%
S_{1}\sim X_{1}$, $S_{2}\sim X_{2}$, and $S_{1}^{\prime }\sim X_{1}^{\prime
} $, $S_{1}^{\prime }\sim X_{1}^{\prime }$.

We represent the head of a coin by $+1$ and its tail by $-1$ and adapt this
convention in the rest of this paper. For coins, Alice's outcome of $\pi
_{A}=+1$ or $-1$ (whether she goes for $S_{1}$ or $S_{2}$) is independent
from Bob's outcome of $\pi _{B}=+1$ or $-1$ (whether he goes for $%
S_{1}^{\prime }$ or $S_{2}^{\prime }$) and relevant joint probabilities are
factorizable.

Referring to the definition (\ref{factorizability (mathematical)}) of
factorizability and noticing that probabilities associated to coins are
factorizable, we use the same notation that is introduced in Section \ref%
{FvBI} to consider, for example, the probability $\Pr (\pi _{A},\pi
_{B};S_{1},S_{1}^{\prime })$ that can be factorized as $\Pr (\pi
_{A};S_{1})\Pr (\pi _{B};S_{1}^{\prime })$.

We define probabilities $r,r^{\prime }\in \lbrack 0,1]$ by $r=\Pr (+1;S_{1})$
and $r^{\prime }=\Pr (+1;S_{1}^{\prime })$ saying that $r$ is the
probability of getting head for Alice's first coin $S_{1}$ and $r^{\prime }$
is the probability of getting head for Bob's first coin $S_{1}^{\prime }$.
Factorizability then allows us to write $\Pr (+1,-1;S_{1},S_{1}^{\prime
})=r(1-r^{\prime })$ and $\Pr (-1,-1;S_{2},S_{2}^{\prime
})=(1-s)(1-s^{\prime })$ where $s=\Pr (+1;S_{2})$ and $s^{\prime }=\Pr
(+1;S_{2}^{\prime })$ i.e. $s$ and $s^{\prime }$ are the probabilities of
getting head for Alice's and Bob's second coin, respectively.

In four-coin setup we find it useful to have the following table:

\begin{equation}
\begin{array}{c}
\text{Alice}%
\end{array}%
\overset{%
\begin{array}{r}
\text{Bob}%
\end{array}%
}{%
\begin{tabular}{c|c}
$%
\begin{array}{c}
S_{1}%
\end{array}%
\begin{array}{c}
+1 \\ 
-1%
\end{array}%
\overset{\overset{%
\begin{array}{c}
S_{1}^{\prime }%
\end{array}%
}{%
\begin{array}{cccccc}
+1 &  &  &  &  & -1%
\end{array}%
}}{\underset{}{%
\begin{tabular}{c|c}
$rr^{\prime }$ & $r(1-r^{\prime })$ \\ \hline
$r^{\prime }(1-r)$ & $(1-r)(1-r^{\prime })$%
\end{tabular}%
\ \ }}$ & $\overset{\overset{%
\begin{array}{c}
S_{2}^{\prime }%
\end{array}%
}{%
\begin{array}{cccccc}
+1 &  &  &  &  & -1%
\end{array}%
}}{\underset{}{%
\begin{tabular}{c|c}
$rs^{\prime }$ & $r(1-s^{\prime })$ \\ \hline
$s^{\prime }(1-r)$ & $(1-r)(1-s^{\prime })$%
\end{tabular}%
\ \ }}$ \\ \hline
$%
\begin{array}{c}
S_{2}%
\end{array}%
\begin{array}{c}
+1 \\ 
-1%
\end{array}%
\overset{}{%
\begin{tabular}{c|c}
$sr^{\prime }$ & $s(1-r^{\prime })$ \\ \hline
$r^{\prime }(1-s)$ & $(1-s)(1-r^{\prime })$%
\end{tabular}%
\ \ }$ & $\overset{}{%
\begin{tabular}{c|c}
$ss^{\prime }$ & $s(1-s^{\prime })$ \\ \hline
$s^{\prime }(1-s)$ & $(1-s)(1-s^{\prime })$%
\end{tabular}%
\ \ }$%
\end{tabular}%
\ \ }  \label{table of factorizeable probabilities}
\end{equation}%
from which we define payoff relations for the players:

\begin{equation}
\begin{array}{c}
\Pi _{A,B}(S_{1},S_{1}^{\prime })=\text{\c{r}}^{T}(\mathcal{A}\mathbf{,}%
\mathcal{B}\mathbf{)}\text{\c{r}}^{\prime },\text{ \ \ }\Pi
_{A,B}(S_{1},S_{2}^{\prime })=\text{\c{r}}^{T}(\mathcal{A}\mathbf{,}\mathcal{%
B}\mathbf{)}\text{\c{s}}^{\prime }, \\ 
\Pi _{A,B}(S_{2},S_{1}^{\prime })=\text{\c{s}}^{T}(\mathcal{A}\mathbf{,}%
\mathcal{B}\mathbf{)}\text{\c{r}}^{\prime }\mathbf{,}\text{ \ \ }\Pi
_{A,B}(S_{2},S_{2}^{\prime })=\text{\c{s}}^{T}(\mathcal{A}\mathbf{,}\mathcal{%
B}\mathbf{)}\text{\c{s}}^{\prime },%
\end{array}
\label{Bilinear payoffs}
\end{equation}%
where

\begin{equation}
\text{\c{r}}=%
\begin{pmatrix}
r \\ 
1-r%
\end{pmatrix}
,\text{ \c{s}}=%
\begin{pmatrix}
s \\ 
1-s%
\end{pmatrix}
,\text{ \c{r}}^{\prime}=%
\begin{pmatrix}
r^{\prime} \\ 
1-r^{\prime}%
\end{pmatrix}
,\text{ \c{s}}^{\prime}=%
\begin{pmatrix}
s^{\prime} \\ 
1-s^{\prime}%
\end{pmatrix}
.
\end{equation}
For example, $\Pi_{A}(S_{1},S_{2}^{\prime})$ is Alice's payoff when, in
repeated runs of coin tossing, she always goes for her first coin, i.e. $%
S_{1}$, while Bob goes for his second coin, i.e. $S_{2}^{\prime}$.

As it is the case with two-coin setup, Alice's mixed strategy in four-coin
setup is the probability with which she chooses her pure strategy\footnote{%
Notice that our definition of a pure strategy corresponds to the usual
mixed-strategy. This agrees with the result in quantum games that a product
pure state corresponds to a mixed-strategy classical game.} $S_{1}$ over her
other pure strategy $S_{2}$ during repeated runs of the experiment.
Similarly, Bob's mixed strategy is the probability with which he chooses his
pure strategy $S_{1}^{\prime }$ over his other pure strategy $S_{2}^{\prime
} $ during repeated runs of the experiment. Assume that Alice plays $S_{1}$
with probability $x$ and Bob plays $S_{1}^{\prime }$ with probability $y$,
their mixed-strategy payoff relations are

\begin{equation}
\Pi _{A,B}(x,y)=%
\begin{pmatrix}
x \\ 
1-x%
\end{pmatrix}%
^{T}%
\begin{pmatrix}
\Pi _{A,B}(S_{1},S_{1}^{\prime }) & \Pi _{A,B}(S_{1},S_{2}^{\prime }) \\ 
\Pi _{A,B}(S_{2},S_{1}^{\prime }) & \Pi _{A,B}(S_{2},S_{2}^{\prime })%
\end{pmatrix}%
\begin{pmatrix}
y \\ 
1-y%
\end{pmatrix}%
.  \label{coins payoff relations}
\end{equation}%
The NE can then be found from (\ref{NE factorable 1}), which is written as

\begin{equation}
\begin{array}{l}
(\text{\c{r}}-\text{\c{s}})^{T}\mathcal{A}\left\{ y^{\star }(\text{\c{r}}%
^{\prime }-\text{\c{s}}^{\prime }\mathbf{)+}\text{\c{s}}^{\prime }\right\}
(x^{\star }-x)\geqslant 0, \\ 
\left\{ x^{\star }(\text{\c{r}}-\text{\c{s}})^{T}+\text{\c{s}}^{T}\right\} 
\mathcal{B}\mathbf{(}\text{\c{r}}^{\prime }\mathbf{-}\text{\c{s}}^{\prime }%
\mathbf{)}(y^{\star }-y)\geqslant 0.%
\end{array}
\label{NE factorable 2}
\end{equation}%
In the following, before we make a transition to playing our game using EPR
experiments, we consider playing three well known games using both the two-
and the four-coin setups.

\subsection{Examples}

We analyze the games of PD, SH, and Chicken in two- and four-coin setups and
afterwards make a transition to the EPR setup. PD is known to be a
representative of the problems of social cooperation \cite{Rasmusen}\ and
has been one of the earliest \cite{EWL} and favorite topics for quantum
games. Hence it is worthwhile to analyze this game in the setup using
non-factorizable probabilities. Our second game is SH that, like PD,
describes conflict between safety and social cooperation. Our third game is
Chicken, also known as the Hawk-Dove game \cite{Rasmusen}, which is
considered an influential model of conflict for two players in game theory.

\subsubsection{\label{PD(C)}Prisoner's Dilemma}

PD is a noncooperative game \cite{Rasmusen} that is widely known to
economists, social and political scientists, and in recent years to quantum
physicists. It is one of the earliest games to be investigated in the
quantum regime \cite{EWL}. Its name comes from the following situation: two
criminals are arrested after having committed a crime together. Each suspect
is placed in a separate cell and may choose between two strategies: \emph{to
confess} $(D)$ and \emph{not to confess} $(C)$, where $C$ and $D$ stand for
Cooperation and Defection.

If neither suspect confesses, i.e. $(C,C)$, they go free, which is
represented by $K$ units of payoff for each suspect. When one prisoner
confesses $(D)$ and the other does not $(C)$, the prisoner who confesses
gets $M$ units of payoff, which represents freedom as well as financial
reward i.e. $M>K$, while the prisoner who did not confess gets $L$,
represented by his ending up in the prison. When both prisoners confess,
i.e. $(D,D)$, both are given a reduced term represented by $N$ units of
payoff, where $N>L$, but it is not so good as going free i.e. $K>N$.

Referring to the matrices (\ref{A and B matrices}) we make the association $%
X_{1},X_{1}^{\prime }\sim $ $C$ and $X_{2},X_{2}^{\prime }\sim $ $D$ and
require that $M>K>N>L$. We define $\triangle _{1}=(M-K),$ $\triangle
_{2}=(N-L),$ $\triangle _{3}=(\triangle _{2}-\triangle _{1})$ which requires 
$\triangle _{1},\triangle _{2}>0$ for this game. In two-coin setup, the
inequalities (\ref{NE factorable 1}) give

\begin{equation}
\begin{array}{l}
\Pi _{A}(x^{\star },y^{\star })-\Pi _{A}(x,y^{\star })=(y^{\star }\triangle
_{3}-\triangle _{2})(x^{\star }-x)\geqslant 0, \\ 
\Pi _{B}(x^{\star },y^{\star })-\Pi _{B}(x^{\star },y)=(x^{\star }\triangle
_{3}-\triangle _{2})(y^{\star }-y)\geqslant 0,%
\end{array}
\label{PD NE 2 coins}
\end{equation}%
and $(x^{\star },y^{\star })=(0,0)$ comes out as a unique NE at which
players' payoffs are $\Pi _{A}(S_{1},S_{1}^{\prime })=N=\Pi
_{B}(S_{1},S_{1}^{\prime })$.

In the four-coin setup, the PD game as defined above is played as follows.
Using the mixed-strategy payoff relation (\ref{coins payoff relations}), the
pair of pure strategies $(S_{2},S_{2}^{\prime })$ is represented by $%
(x^{\star },y^{\star })=(0,0)$. If we require this strategy pair to be a NE
then we also need to know about the constraints this requirement imposes on $%
r,s,r^{\prime },s^{\prime }$. When $(x^{\star },y^{\star })=(0,0)$ the NE
inequalities (\ref{NE factorable 2}) for PD are reduced to

\begin{equation}
\begin{array}{l}
-x(s-r)\triangle _{2}\left\{ \left( \triangle _{1}/\triangle _{2}-1\right)
s+1\right\} \geqslant 0, \\ 
-y(s^{\prime }-r^{\prime })\triangle _{2}\left\{ \left( \triangle
_{1}/\triangle _{2}-1\right) s^{\prime }+1\right\} \geqslant 0.%
\end{array}
\label{NE Generalized PD}
\end{equation}%
Now, for the NE inequalities (\ref{NE Generalized PD}) to hold, it is
required that $(s-r)\leq 0$ and $(s^{\prime }-r^{\prime })\leq 0$ both when $%
\triangle _{1}/\triangle _{2}\geq 1$ and when $\triangle _{1}/\triangle
_{2}<1$. This, of course, is possible if

\begin{equation}
\Pr \{S_{2}(+1)\}=s=0=s^{\prime }=\Pr \{S_{2}^{\prime }(+1)\},
\label{s=0=s'}
\end{equation}%
which must hold if the strategy pair $(x^{\star },y^{\star })=(0,0)$ is to
be a NE in PD. Eqs. (\ref{s=0=s'}) should be true along with that the
probabilities $\Pr (\pi _{A},\pi _{B};a,b)$ are factorizable. As we find it,
this result provides the basis on which the forthcoming argument for the
quantum version of this game rests. Notice that, from (\ref{Bilinear payoffs}%
), we obtain $\Pi _{A}(S_{2},S_{2}^{\prime })=\Pi _{B}(S_{2},S_{2}^{\prime
})=N$ when Eq. (\ref{s=0=s'}) holds.

The constraint (\ref{s=0=s'}) appears when the strategy pair $%
(S_{2},S_{2}^{\prime})$ is assumed to be the NE. One can assume other
strategy pair, for example $(S_{1},S_{1}^{\prime})$, to be a NE for which,
instead of the requirement (\ref{s=0=s'}), we obtain

\begin{equation}
\Pr \{S_{1}(+1)\}=r=0=r^{\prime }=\Pr \{S_{1}^{\prime }(+1)\}.
\end{equation}%
However, it is found that this freedom does not affect the forthcoming
argument for a quantum game.

\subsubsection{\label{SH}Stag Hunt}

Along with PD, the game of SH provides another interesting context to study
problems of social cooperation. It describes the situation when two hunters
can either jointly hunt a stag (an adult deer that makes a large meal) or
individually hunt a rabbit (which is tasty but makes substantially small
meal). Hunting a stag is quite challenging and hunters need to cooperate
with each other, especially, it is quite unlikely that a hunter hunts a stag
alone.

It is found that, in contrast to PD that has a single pure NE, the game of
SH has three NE, two of which are pure and one is mixed. The two pure NE
correspond to the situations when both hunters hunt the stag as a team and
when each hunts rabbit by himself. SH differs from PD in that mutual
cooperation gives maximum reward to the hunters. When compared to PD, SH is
considered a better model for the problems of (social) cooperation.

Referring to the matrices (\ref{A and B matrices}) the game of SH is defined
by

\begin{equation}
K>M\geq N>L\text{ and }M+N>K+L.  \label{Stag hunt}
\end{equation}
In two-coin setup the NE inequalities for this game remain the same as the
inequalities (\ref{PD NE 2 coins}) except that now we have

\begin{equation}
\triangle _{3}>\triangle _{2}>0\text{ and }0>\triangle _{1},
\label{Stag Hunt condition}
\end{equation}%
instead of the condition $\triangle _{1},\triangle _{2}>0$ that hold for PD.
Here $\triangle _{1},\triangle _{2},\triangle _{3}$ are defined in the
Section \ref{PD(C)}. This leads to three NE:

\begin{equation}
\begin{array}{l}
(x^{\star },y^{\star })_{1}=(0,0); \\ 
(x^{\star },y^{\star })_{2}=(\triangle _{2}/\triangle _{3},\triangle
_{2}/\triangle _{3}); \\ 
(x^{\star },y^{\star })_{3}=(1,1);%
\end{array}
\label{SH 2 coin three NE}
\end{equation}%
and the corresponding payoffs at these equilibria, obtained from Eqs. (\ref%
{2-Coin payoff relations}), are

\begin{equation}
\begin{array}{l}
\Pi _{A}(x^{\star },y^{\star })_{1}=N=\Pi _{B}(x^{\star },y^{\star })_{1};
\\ 
\Pi _{A}(x^{\star },y^{\star })_{2}=(\triangle _{2}/\triangle
_{3})^{2}\triangle _{3}+(\triangle _{2}/\triangle _{3})\triangle _{4}+N=\Pi
_{B}(x^{\star },y^{\star })_{2}; \\ 
\Pi _{A}(x^{\star },y^{\star })_{3}=K=\Pi _{B}(x^{\star },y^{\star })_{3};%
\end{array}
\label{SH payoffs}
\end{equation}%
where we define $\triangle _{4}=L+M-2N$.

Now consider playing this game within the four-coin setup in which the NE
inequalities (\ref{NE factorable 2}) reduce to

\begin{equation}
\begin{array}{l}
(r-s)[y^{\star }(r^{\prime }-s^{\prime })\triangle _{3}+(s^{\prime
}\triangle _{3}-\triangle _{2})](x^{\star }-x)\geq 0, \\ 
(r^{\prime }-s^{\prime })[x^{\star }(r-s)\triangle _{3}+(s\triangle
_{3}-\triangle _{2})](y^{\star }-y)\geq 0.%
\end{array}
\label{NE Stag hunt}
\end{equation}%
From the inequalities (\ref{NE Stag hunt}) the NE $(x^{\star },y^{\star
})_{1}=(0,0)$ results when

\begin{equation}
s=0=s^{\prime}  \label{SH pure 1}
\end{equation}
and, similarly, the NE $(x^{\star},y^{\star})_{3}=(1,1)$ results when

\begin{equation}
r=0=r^{\prime}.  \label{SH pure 2}
\end{equation}
Also, the inequalities (\ref{NE Stag hunt}) hold when $x^{\star}=(s%
\triangle_{3}-\triangle_{2})/(s-r)\triangle_{3}$, $y^{\star}=(s^{\prime
}\triangle_{3}-\triangle_{2})/(s^{\prime}-r^{\prime})\triangle_{3}$ and for $%
(x^{\star},y^{\star})_{2}=(\triangle_{2}/\triangle_{3},\triangle
_{2}/\triangle_{3})$ to be a NE we require

\begin{equation}
s=0\text{, }r=1\text{ and }s^{\prime}=0\text{, }r^{\prime}=1.
\label{SH probs constraints}
\end{equation}
These constraints on $r,s,r^{\prime},s^{\prime}$ hold along with the
probabilities $\Pr(\pi_{A},\pi_{B};a,b)$ being factorizable.

\subsubsection{\label{CG}Chicken game}

The game of Chicken is about two drivers who drive towards each other from
opposite directions. One driver must turn aside, or both may die in a crash.
If one driver turns aside but the other does not, s/he will be called a
\textquotedblleft chicken\textquotedblright . While each driver prefers not
to yield to the opponent, the outcome where neither driver yields is the
worst possible one for both. In this anti-coordination game it is mutually
beneficial for parties to play different strategies.

Sometimes, Chicken is also known as the \textquotedblleft
Hawk-Dove\textquotedblright\ game, that originates from the parallel
development of the basic principles of this game in two different research
areas: economics and mathematical biology. Economists, and the political
scientists too, refer to \cite{Aumann} this game as Chicken,\ while
mathematical biologists refer to \cite{Hofbauer Sigmund} it as the
Hawk-Dove\ game.

The game of Chicken differs from PD in that in Chicken the mutual defection
(the crash when both players drive straight) is the most feared outcome.
While in PD cooperation while the other player defects is the worst outcome.

A version of the Chicken game is obtained from the matrices (\ref{A and B
matrices}) when

\begin{equation}
K=0,\text{ }L=\alpha ,\text{ }M=\beta ,\text{ }N=0,\text{ }0<\alpha <(\alpha
+\beta ).  \label{chicken game}
\end{equation}%
Playing this game in two-coin setup the inequalities (\ref{NE factorable 1})
are reduced to

\begin{equation}
\left\{ \alpha -y^{\star }(\alpha +\beta )\right\} (x^{\star }-x)\geq 0,%
\text{ \ \ }\left\{ \alpha -x^{\star }(\alpha +\beta )\right\} (y^{\star
}-y)\geq 0,  \label{Chicken 2 coin NE}
\end{equation}%
and three NE emerge:

\begin{equation}
\begin{array}{l}
(x^{\star },y^{\star })_{1}=(1,0); \\ 
(x^{\star },y^{\star })_{2}=(\alpha /(\alpha +\beta ),\alpha /(\alpha +\beta
)); \\ 
(x^{\star },y^{\star })_{3}=(0,1).%
\end{array}
\label{3 Equilibria in Chicken}
\end{equation}%
The corresponding payoffs at these equilibria, obtained from Eqs. (\ref%
{2-Coin payoff relations}), are

\begin{equation}
\begin{array}{l}
\Pi _{A}(x^{\star },y^{\star })_{1}=\alpha ,\text{ \ \ }\Pi _{B}(x^{\star
},y^{\star })_{1}=\beta ; \\ 
\Pi _{A}(x^{\star },y^{\star })_{2}=\alpha \beta /(\alpha +\beta )=\Pi
_{B}(x^{\star },y^{\star })_{2}; \\ 
\Pi _{A}(x^{\star },y^{\star })_{3}=\beta ,\text{ \ \ }\Pi _{B}(x^{\star
},y^{\star })_{3}=\alpha .%
\end{array}
\label{CG NE payoffs}
\end{equation}

Now we play this game using the four-coin setup. The NE inequalities come
out to be the same as the ones given in (\ref{NE Stag hunt}) except that now
we have $\triangle _{3}=-(\alpha +\beta )$ and $\triangle _{2}=-\alpha $.
Then for $(x^{\star },y^{\star })_{1}=(1,0)$ we require

\begin{equation}
r=0\text{ and }s^{\prime}=0.  \label{CG pure 1}
\end{equation}
Similarly, for $(x^{\star},y^{\star})_{3}=(0,1)$ we require

\begin{equation}
r^{\prime}=0\text{ and }s=0.  \label{CG pure 2}
\end{equation}
At $(x^{\star},y^{\star})_{2}=(\alpha/(\alpha+\beta),\alpha/(\alpha+\beta))$
the inequalities (\ref{NE Stag hunt}) reduce to

\begin{equation}
(r-s)(\alpha -\alpha r^{\prime }-\beta s^{\prime })(\alpha /(\alpha +\beta
)-x)\geq 0,\text{ \ \ }(r^{\prime }-s^{\prime })(\alpha -\alpha r-\beta
s)(\alpha /(\alpha +\beta )-y)\geq 0,
\end{equation}%
which puts constraint on $r,s,r^{\prime },s^{\prime }$ given as

\begin{equation}
\alpha(1-r^{\prime})=\beta s^{\prime},\text{ \ \ }\alpha(1-r)=\beta s.
\label{Chicken game constraints on r,s,r',s'}
\end{equation}

A special case is the one when $\alpha=\beta$ and the strategy pair $%
(x^{\star},y^{\star})=(1/2,1/2)$ becomes a NE which imposes certain
constraints on $r,s,r^{\prime},s^{\prime}$. For this NE the inequalities (%
\ref{NE factorable 2}), for the game defined by (\ref{A and B matrices}, \ref%
{chicken game}), are reduced to

\begin{equation}
\begin{array}{l}
(r-s)\left\{ -(\alpha +\beta )(r^{\prime }+s^{\prime })/2+L\right\}
(1/2-x)\geqslant 0, \\ 
(r^{\prime }-s^{\prime })\left\{ -(\alpha +\beta )(r+s)/2+L\right\}
(1/2-y)\geqslant 0,%
\end{array}
\label{Chicken NE}
\end{equation}%
so, we require

\begin{equation}
r+s=1=r^{\prime }+s^{\prime },  \label{Chicken constraint}
\end{equation}%
if the strategy pair $(x^{\star },y^{\star })=(1/2,1/2)$ is to be a NE in
this game. Along with this, the probabilities $\Pr (\pi _{A},\pi _{B};a,b)$
are to be factorizable.

\section{\label{tpguEPRs}Playing games with EPR experiments}

Section \ref{tpgFPs} describes playing a two-player game with four coins
such that choosing a coin is a strategy while players' payoffs are given by
their strategies, the matrix of the game, and the underlying statistics of
the coins. This facilitates transition to playing the \emph{same} game using
EPR experiments.

In EPR setup, Alice and Bob are spatially separated and are unable to
communicate with each other. In an individual run, both receive one half of
a pair of particles originating from a common source. In the same run of the
experiment both choose one from two given (pure) strategies. These
strategies are the two directions in space along which spin or polarization
measurements can be made.

Keeping the notation for the coins, we denote these directions to be $S_{1}$%
, $S_{2}$ for Alice and $S_{1}^{\prime }$, $S_{2}^{\prime }$ for Bob. Each
measurement generates $+1$\ or $-1$\ as the outcome, like it is the case
with coins after their toss in the four-coin setup. Experimental results are
recorded for a large number of individual runs of the experiment and payoffs
are awarded depending on the directions the players go for over many runs
(defining their strategies), the matrix of the game they play, and the
statistics of the measurement outcomes.

For EPR experiments, we retain Cereceda' notation \cite{Cereceda} for the
associated probabilities: 
\begin{equation}
p_{k}=\Pr (\pi _{A},\pi _{B};a,b){\ \ \ }\text{with}{\mathrm{\ \ \ }}k=1+%
\frac{(1-\pi _{B})}{2}+2\frac{(1-\pi _{A})}{2}+4(b-1)+8(a-1).
\label{Cerenotation}
\end{equation}%
In this notation, for example, we write $p_{1}$ for the probability $\Pr
(+1,+1;S_{1},S_{1}^{\prime })$ and $p_{8}$ for the probability $\Pr
(-1,-1;S_{1},S_{2}^{\prime })$. One can then construct the following table
of probabilities

\begin{equation}
\begin{array}{c}
\text{Alice}%
\end{array}%
\begin{array}{c}
\underset{}{%
\begin{array}{c}
S_{1}%
\end{array}%
\begin{array}{c}
+1 \\ 
-1%
\end{array}%
} \\ 
\overset{}{%
\begin{array}{c}
S_{2}%
\end{array}%
\begin{array}{c}
+1 \\ 
-1%
\end{array}%
}%
\end{array}%
\overset{\overset{%
\begin{array}{c}
\text{Bob}%
\end{array}%
}{%
\begin{array}{cc}
\overset{%
\begin{array}{c}
S_{1}^{\prime }%
\end{array}%
}{%
\begin{array}{cc}
+1 & -1%
\end{array}%
} & \overset{%
\begin{array}{c}
S_{2}^{\prime }%
\end{array}%
}{%
\begin{array}{cc}
+1 & -1%
\end{array}%
}%
\end{array}%
}}{%
\begin{tabular}{c|c}
$\underset{}{%
\begin{array}{cc}
p_{1} & p_{2} \\ 
p_{3} & p_{4}%
\end{array}%
}$ & $\underset{}{%
\begin{array}{cc}
p_{5} & p_{6} \\ 
p_{7} & p_{8}%
\end{array}%
}$ \\ \hline
$\overset{}{%
\begin{array}{cc}
p_{9} & p_{10} \\ 
p_{11} & p_{12}%
\end{array}%
}$ & $\overset{}{%
\begin{array}{cc}
p_{13} & p_{14} \\ 
p_{15} & p_{16}%
\end{array}%
}$%
\end{tabular}%
}.  \label{EPR probabilities}
\end{equation}%
This table allows to transparently see how the probabilities $p_{i}(1\leq
i\leq 16)$ are linked to the probabilities $\Pr (\pi _{A},\pi _{B};a,b)$,
where we recall that $a$ can be set at $S_{1}$ or at $S_{2}$ and, similarly, 
$b$ can be set at $S_{2}$ or at $S_{2}^{\prime }$. In Cereceda's notation
the EPR probabilities $p_{i}$ are normalized as they satisfy the following
relations

\begin{equation}
\begin{array}{l}
p_{1}+p_{2}+p_{3}+p_{4}=1, \\ 
p_{5}+p_{6}+p_{7}+p_{8}=1, \\ 
p_{9}+p_{10}+p_{11}+p_{12}=1, \\ 
p_{13}+p_{14}+p_{15}+p_{16}=1.%
\end{array}
\label{Normalization}
\end{equation}%
Notice that the factorizable probabilities (\ref{table of factorizeable
probabilities}) are also normalized and (\ref{Normalization}) holds for them.

Payoff relations (\ref{Bilinear payoffs}) are originally constructed when
the game given by the matrices (\ref{A and B matrices}) is played with four
coins and their mathematical form convinces one to use the following recipe 
\cite{Iqbal(1)} to reward players when the same game is played using EPR
probabilities (\ref{EPR probabilities}):

\begin{equation}
\begin{array}{l}
\Pi _{A}(S_{1},S_{1}^{\prime })=Kp_{1}+Lp_{2}+Mp_{3}+Np_{4}, \\ 
\Pi _{A}(S_{1},S_{2}^{\prime })=Kp_{5}+Lp_{6}+Mp_{7}+Np_{8}, \\ 
\Pi _{A}(S_{2},S_{1}^{\prime })=Kp_{9}+Lp_{10}+Mp_{11}+Np_{12}, \\ 
\Pi _{A}(S_{2},S_{2}^{\prime })=Kp_{13}+Lp_{14}+Mp_{15}+Np_{16}.%
\end{array}
\label{EPR Payoffs}
\end{equation}%
Here $\Pi _{A}(S_{1},S_{2}^{\prime })$, for example, is Alice's payoff when
she plays $S_{1}$ and Bob plays $S_{2}^{\prime }$. Like it is the case with
four coins, the payoff relations for Bob are obtained from (\ref{EPR Payoffs}%
) by the transformation $L\leftrightarrows M$ in Eqs. (\ref{EPR Payoffs}).

When $p_{i}$ are factorizable in terms of $r,r^{\prime },s,s^{\prime }$, a
comparison of (\ref{EPR Payoffs}) with (\ref{Bilinear payoffs}) requires

\begin{equation}
\begin{array}{llll}
p_{1}=rr^{\prime }, & p_{2}=r(1-r^{\prime }), & p_{3}=r^{\prime }(1-r), & 
p_{4}=(1-r)(1-r^{\prime }), \\ 
p_{5}=rs^{\prime }, & p_{6}=r(1-s^{\prime }), & p_{7}=s^{\prime }(1-r), & 
p_{8}=(1-r)(1-s^{\prime }), \\ 
p_{9}=sr^{\prime }, & p_{10}=s(1-r^{\prime }), & p_{11}=r^{\prime }(1-s), & 
p_{12}=(1-s)(1-r^{\prime }), \\ 
p_{13}=ss^{\prime }, & \text{\ }p_{14}=s(1-s^{\prime }), & p_{15}=s^{\prime
}(1-s), & p_{16}=(1-s)(1-s^{\prime }).%
\end{array}
\label{Pis written bilinearly}
\end{equation}%
That is, the factorizability of $p_{i}$ in terms $r,r^{\prime },s,s^{\prime }
$ makes the game played by EPR probabilities equivalent to the one played by
using coins.

However, the EPR probabilities $p_{i}$, appearing in (\ref{Bilinear payoffs}%
), may not be factorizable in terms of $r,s,r^{\prime},s^{\prime}$, whereas
for both the payoff relations (\ref{Bilinear payoffs}, \ref{EPR Payoffs})
the normalization (\ref{Normalization}) continues to hold.

\section{\label{tpgunfps}Two-player games using non-factorizable
probabilities}

As it is the case with the coin game, Alice's mixed strategy is defined to
be the probability to choose between $S_{1}$ and $S_{2}$ and we can use,
once again, the payoff relations (\ref{coins payoff relations}) which,
however, now correspond to the possible situation when $p_{i}$ may not be
factorizable. So that, the relations (\ref{Bilinear payoffs}) can be
replaced with the relations (\ref{EPR Payoffs}) in Alice's mixed-strategy
payoff relation in (\ref{coins payoff relations}). The same applies to Bob's
payoff relations.

Note that when $p_{i}$ are factorizable, using (\ref{Pis written bilinearly}%
) allows the probabilities $r,r^{\prime},s,s^{\prime}$ to be expressed in
terms of $p_{i}$:

\begin{equation}
r=p_{1}+p_{2},\text{ \ \ }s=p_{9}+p_{10},\text{ \ \ }r^{\prime }=p_{1}+p_{3},%
\text{ \ \ }s^{\prime }=p_{5}+p_{7},  \label{discrete prob}
\end{equation}%
which are useful relations for the forthcoming argument for a quantum game.

Along with the normalization (\ref{Normalization}), the EPR probabilities $%
p_{i}$ $(1\leq i\leq 16)$ also satisfy certain other constraints imposed by
the requirements of causality. Cereceda \cite{Cereceda} writes these
constraints as

\begin{equation}
\begin{array}{ll}
p_{1}+p_{2}-p_{5}-p_{6}=0, & p_{1}+p_{3}-p_{9}-p_{11}=0, \\ 
p_{9}+p_{10}-p_{13}-p_{14}=0, & p_{5}+p_{7}-p_{13}-p_{15}=0, \\ 
p_{3}+p_{4}-p_{7}-p_{8}=0, & p_{11}+p_{12}-p_{15}-p_{16}=0, \\ 
p_{2}+p_{4}-p_{10}-p_{12}=0, & p_{6}+p_{8}-p_{14}-p_{16}=0,%
\end{array}
\label{Constraints on Ps}
\end{equation}%
which is referred to as the \emph{causal communication constraint }\cite%
{Cereceda}. Notice that the constraints (\ref{Constraints on Ps}), of
course, also hold when $p_{i}$ are factorizable and are written in terms of $%
r,s,r^{\prime },s^{\prime }$ as in (\ref{Pis written bilinearly}).
Essentially, these constraints state that, on measurement, Alice's
probability of obtaining particular outcome ($+1$ or $-1$), when she goes
for $S_{1}$ or $S_{2}$, is independent of how Bob sets up his apparatus
(i.e. along $S_{1}^{\prime }$ or along $S_{2}^{\prime }$). The same applies
to Bob i.e. on measurement his probability of obtaining a particular outcome
($+1$ or $-1$), when he goes for $S_{1}^{\prime }$ or $S_{2}^{\prime }$, is
independent of how Alice sets up her apparatus (i.e. along $S_{1}$ or along $%
S_{2}$). Other authors may like to call the constraints (\ref{Constraints on
Ps}) with some different name, for example, Winsberg and Fine \cite%
{WinsbergFine}\ have described them as the \emph{locality constraint}.

Notice that because of normalization (\ref{Normalization}) half of the Eqs. (%
\ref{Constraints on Ps}) are redundant that makes eight among sixteen
probabilities $p_{i}$ to be independent. A convenient solution \cite%
{Cereceda} of the system (\ref{Normalization}, \ref{Constraints on Ps}), for
which the set of variables:

\begin{equation}
\upsilon =\left\{
p_{2},p_{3},p_{6},p_{7},p_{10},p_{11},p_{13},p_{16}\right\} ,
\label{First set of probabilities}
\end{equation}%
is expressed in terms of the remaining set of variables:

\begin{equation}
\mu=\left\{ p_{1},p_{4},p_{5},p_{8},p_{9},p_{12},p_{14},p_{15}\right\} ,
\label{Second set of probabilities}
\end{equation}
is given as follows

\begin{equation}
\begin{array}{l}
p_{2}=(1-p_{1}-p_{4}+p_{5}-p_{8}-p_{9}+p_{12}+p_{14}-p_{15})/2, \\ 
p_{3}=(1-p_{1}-p_{4}-p_{5}+p_{8}+p_{9}-p_{12}-p_{14}+p_{15})/2, \\ 
p_{6}=(1+p_{1}-p_{4}-p_{5}-p_{8}-p_{9}+p_{12}+p_{14}-p_{15})/2, \\ 
p_{7}=(1-p_{1}+p_{4}-p_{5}-p_{8}+p_{9}-p_{12}-p_{14}+p_{15})/2, \\ 
p_{10}=(1-p_{1}+p_{4}+p_{5}-p_{8}-p_{9}-p_{12}+p_{14}-p_{15})/2, \\ 
p_{11}=(1+p_{1}-p_{4}-p_{5}+p_{8}-p_{9}-p_{12}-p_{14}+p_{15})/2, \\ 
p_{13}=(1-p_{1}+p_{4}+p_{5}-p_{8}+p_{9}-p_{12}-p_{14}-p_{15})/2, \\ 
p_{16}=(1+p_{1}-p_{4}-p_{5}+p_{8}-p_{9}+p_{12}-p_{14}-p_{15})/2.%
\end{array}
\label{dependent probabilities}
\end{equation}%
The relationships (\ref{dependent probabilities}) between joint
probabilities arise because both the normalization condition (\ref%
{Normalization}) and the causal communication constraint (\ref{Constraints
on Ps}) are fulfilled.

From Eqs. (\ref{dependent probabilities}) one can obtain other constraints
considering that the sum of any combination of probabilities from the set $%
\upsilon$ must be non-negative. In the following are some results to be used
later in this paper. In (\ref{dependent probabilities}) the sum $p_{2}+p_{7}$
is non-negative and it requires that

\begin{equation}
p_{1}+p_{8}\leq1.  \label{p1 and p8}
\end{equation}
In (\ref{dependent probabilities}) the sum $p_{3}+p_{10}$ is non-negative
and it requires that

\begin{equation}
p_{1}+p_{12}\leq1.  \label{p1 and p12}
\end{equation}
Similarly, the sum $p_{6}+p_{13}$ is non-negative and it requires that

\begin{equation}
p_{8}+p_{13}\leq1.  \label{p8 and p13}
\end{equation}
Both of the inequalities (\ref{p1 and p8}, \ref{p1 and p12}) are found
useful in developing a quantum version of PD. The inequality (\ref{p8 and
p13}) is found useful in developing quantum version of SH.

Using (\ref{EPR Payoffs}) in (\ref{coins payoff relations}), with the
assumption that $(x^{\star },y^{\star })$ is a NE, one obtains

\begin{equation}
\begin{array}{l}
\Pi _{A}(x^{\star },y^{\star })-\Pi _{A}(x,y^{\star })=(x^{\star
}-x)[y^{\star }\left\{ K\Omega _{1}+L\Omega _{2}+M\Omega _{3}+N\Omega
_{4}\right\} + \\ 
\left\{ K\left( p_{5}-p_{13}\right) +L\left( p_{6}-p_{14}\right) +M\left(
p_{7}-p_{15}\right) +N\left( p_{8}-p_{16}\right) \right\} ]\geq 0,%
\end{array}
\label{NEx1}
\end{equation}%
and

\begin{equation}
\begin{array}{l}
\Pi _{B}(x^{\star },y^{\star })-\Pi _{B}(x^{\star },y)=(y^{\star
}-y)[x^{\star }\left\{ K\Omega _{1}+M\Omega _{2}+L\Omega _{3}+N\Omega
_{4}\right\} + \\ 
\left\{ K\left( p_{9}-p_{13}\right) +M\left( p_{10}-p_{14}\right) +L\left(
p_{11}-p_{15}\right) +N\left( p_{12}-p_{16}\right) \right\} ]\geq 0,%
\end{array}
\label{NEy1}
\end{equation}%
where

\begin{equation}
\begin{array}{l}
\Omega _{1}=p_{1}-p_{5}-p_{9}+p_{13},\text{ \ \ }\Omega
_{2}=p_{2}-p_{6}-p_{10}+p_{14}, \\ 
\Omega _{3}=p_{3}-p_{7}-p_{11}+p_{15},\text{ \ \ }\Omega
_{4}=p_{4}-p_{8}-p_{12}+p_{16}.%
\end{array}
\label{Omega Definitions}
\end{equation}%
Now use (\ref{dependent probabilities}) to write (\ref{NEx1}) and (\ref{NEy1}%
) in terms of the probabilities appearing in the set $\mu $ given in (\ref%
{Second set of probabilities}) to obtain

\begin{equation}
\begin{array}{l}
\Pi _{A}(x^{\star },y^{\star })-\Pi _{A}(x,y^{\star })=(1/2)(x^{\star
}-x)[y^{\star }\triangle _{3}\times  \\ 
(1+p_{1}+p_{4}-p_{5}-p_{8}-p_{9}-p_{12}-p_{14}-p_{15})- \\ 
\{\triangle _{3}(1-p_{5}-p_{8}-p_{14}-p_{15})+ \\ 
(\triangle _{1}+\triangle _{2})(p_{1}-p_{4}-p_{9}+p_{12})\}]\geq 0,%
\end{array}
\label{NEx}
\end{equation}

\begin{equation}
\begin{array}{l}
\Pi _{B}(x^{\star },y^{\star })-\Pi _{B}(x^{\star },y)=(1/2)(y^{\star
}-y)[x^{\star }\triangle _{3}\times  \\ 
(1+p_{1}+p_{4}-p_{5}-p_{8}-p_{9}-p_{12}-p_{14}-p_{15})- \\ 
\{\triangle _{3}(1-p_{9}-p_{12}-p_{14}-p_{15})+ \\ 
(\triangle _{1}+\triangle _{2})(p_{1}-p_{4}-p_{5}+p_{8})\}]\geq 0.%
\end{array}
\label{NEy}
\end{equation}%
Notice that the probabilities associated to the EPR experiments can be
factorized only for certain directions of measurements even for singlet
states. For these directions the game played using EPR experiments can thus
be interpreted within the four-coin setup.

Essentially, we obtain quantum game from the classical as follows. Referring
to the four-coin setup developed in the Section \ref{tpgFPs}, the
factorizability of associated probabilities in terms of $r,s,r^{\prime
},s^{\prime}$ allows us to translate the requirement that the resulting game
has a classical interpretation into certain constraints on $r,s,r^{\prime
},s^{\prime}$. We find that from factorizability the relations (\ref%
{discrete prob}) follow and from these relations the constraints on $%
r,s,r^{\prime},s^{\prime}$ can be re-expressed in terms of $p_{i}$ $(1\leq
i\leq16)$. We now obtain a quantum version of the game by retaining these
constraints and afterwards allowing $p_{i}$ to become non-factorizable. In
this procedure retaining the constraints ensures that classical outcome
results when probabilities become factorizable.

\subsection{Examples}

In the following we consider the impact of non-factorizable probabilities on
the NE in PD, SH, and the Chicken game.

\subsubsection{\label{QPD}Prisoner's Dilemma}

Recall that Section \ref{tpgFPs} states the result that when PD is played
with four coins we require the condition (\ref{s=0=s'}) to hold if the
strategy pair $(S_{2},S_{2}^{\prime})$ is to exist as a NE. Along with this
the probabilities $p_{i}$ are to be factorizable.

This motivates to construct a quantum version of PD when probabilities $%
p_{i} $ are not factorizable while the constraint (\ref{s=0=s'}) remains
valid. The condition (\ref{s=0=s'}) ensures that with factorizable
probabilities the game can be interpreted classically.

Notice that when the probabilities $p_{i}$ are factorizable, i.e. they can
be written as in (\ref{Pis written bilinearly}), the constraint (\ref{s=0=s'}%
) can hold when numerical values are assigned to certain probabilities among 
$p_{i}$:

\begin{equation}
p_{5}=0,\text{ }p_{7}=0,\text{ }p_{9}=0,\text{ }p_{10}=0,\text{ }p_{16}=1,
\label{probability value assignment}
\end{equation}%
where, because of the normalization (\ref{Normalization}), $p_{16}=1$
requires that $p_{13}=0,$ $p_{14}=0,$ and $p_{15}=0$. This can also be
noticed more directly from (\ref{discrete prob}). This assignment of values
to certain probabilities reduces Eqs. (\ref{Normalization}) and Eqs. (\ref%
{Constraints on Ps}) to

\begin{equation}
\begin{array}{l}
p_{1}+p_{2}+p_{3}+p_{4}=1, \\ 
p_{1}+p_{2}=p_{6},\text{ \ \ \ \ }p_{1}+p_{3}=p_{11}, \\ 
p_{3}+p_{4}=p_{8},\text{ \ \ \ \ }p_{11}+p_{12}=1, \\ 
p_{2}+p_{4}=p_{12},\text{ \ \ \ \ }p_{6}+p_{8}=1.%
\end{array}
\label{QPD Cond2}
\end{equation}%
Substituting from (\ref{probability value assignment}) into (\ref{NEx}, \ref%
{NEy}) gives

\begin{equation}
\begin{array}{l}
\Pi _{A}(x^{\star },y^{\star })-\Pi _{A}(x,y^{\star }) \\ 
=(1/2)(x^{\star }-x)[y^{\star }\triangle _{3}(1+p_{1}+p_{4}-p_{8}-p_{12})-
\\ 
\{\triangle _{3}(1-p_{8})+(\triangle _{1}+\triangle
_{2})(p_{1}-p_{4}+p_{12})\}]\geq 0;%
\end{array}
\label{NE(a)1}
\end{equation}

\begin{equation}
\begin{array}{l}
\Pi _{B}(x^{\star },y^{\star })-\Pi _{B}(x^{\star },y) \\ 
=(1/2)(y^{\star }-y)[x^{\star }\triangle _{3}(1+p_{1}+p_{4}-p_{8}-p_{12})-
\\ 
\{\triangle _{3}(1-p_{12})+(\triangle _{1}+\triangle
_{2})(p_{1}-p_{4}+p_{8})\}]\geq 0.%
\end{array}
\label{NE(b)1}
\end{equation}

Note that from (\ref{QPD Cond2}) we obtain $(1-p_{8})=p_{6}=p_{1}+p_{2}$,
that for factorizable probabilities, becomes equal to $r$ when we refer to
Eqs. (\ref{discrete prob}). Similarly, from (\ref{QPD Cond2}) we obtain $%
(p_{1}-p_{4}+p_{12})=p_{1}+p_{2}=r$. Substituting these in (\ref{NE(a)1})
along with that $x^{\star}=0=y^{\star}$ gives $\Pi_{A}(0,0)-\Pi_{A}(x,0)=xr%
\triangle_{2}\geq0$. In similar way we find from (\ref{QPD Cond2}) that $%
(1-p_{12})=p_{11}=p_{1}+p_{3}$ which for factorizable probabilities becomes
equal to $r^{\prime}$ when we use Eqs. (\ref{discrete prob}). Likewise, from
(\ref{QPD Cond2}) we obtain $(p_{1}-p_{4}+p_{8})=p_{1}+p_{3}=r^{\prime}$.
Substituting these in (\ref{NE(b)1}), along with that $x^{\star}=0=y^{\star}$%
, gives $\Pi_{B}(0,0)-\Pi_{B}(0,y)=yr^{\prime}\triangle_{2}\geq0$. This can
be described as follows: When probabilities $p_{1},p_{4},p_{8}$ and $p_{12}$
are factorizable and the values assigned to them in (\ref{probability value
assignment}) hold, the inequalities (\ref{NE(a)1}, \ref{NE(b)1}) ensure that
the strategy pair $(S_{2},S_{2}^{\prime})$ becomes a NE.

Now we ask about the fate of the NE\ strategy pair $(S_{2},S_{2}^{\prime })$
when in (\ref{probability value assignment}) the values assigned to certain
probabilities, resulting from the requirement (\ref{s=0=s'}), hold while $%
p_{i}$ do not remain factorizable in terms of $r,s,r^{\prime },s^{\prime }$.
Allow the probabilities $p_{1},p_{4},p_{8}$ and $p_{12}$ not to be
factorizable and use (\ref{probability value assignment}) in (\ref{dependent
probabilities}) to get $1-p_{1}+p_{4}-p_{8}-p_{12}=0$ and the inequalities (%
\ref{NE(a)1}, \ref{NE(b)1}) take the form:

\begin{equation}
\begin{array}{l}
\Pi _{A}(x^{\star },y^{\star })-\Pi _{A}(x,y^{\star }) \\ 
=(x^{\star }-x)\left[ \triangle _{2}\left\{ y^{\star
}-(1-p_{8})/p_{1}\right\} -\triangle _{1}y^{\star }\right] p_{1}\geq 0,%
\end{array}
\label{QPD Q NE 1}
\end{equation}

\begin{equation}
\begin{array}{l}
\Pi _{B}(x^{\star },y^{\star })-\Pi _{B}(x^{\star },y) \\ 
=(y^{\star }-y)\left[ \triangle _{2}\left\{ x^{\star
}-(1-p_{12})/p_{1}\right\} -\triangle _{1}x^{\star }\right] p_{1}\geq 0,%
\end{array}
\label{QPD Q NE 2}
\end{equation}%
where $\triangle _{1}$ and $\triangle _{2}$ are defined in the Section (\ref%
{PD(C)}). Note that (\ref{p1 and p8}) and (\ref{p1 and p12}) give $1\leq
(1-p_{8})/p_{1}$ and $1\leq (1-p_{12})/p_{1}$ so that

\begin{equation}
\left\{ y^{\star }-(1-p_{8})/p_{1}\right\} \leq 0,\text{ \ \ }\left\{
x^{\star }-(1-p_{12})/p_{1}\right\} \leq 0,  \label{QPD Q NE 3}
\end{equation}%
which results, once again, in the strategy pair $(x^{\star },y^{\star
})=(0,0)$ being a NE, which is the classical outcome of the game.

Notice that this NE emerges for non-factorizable EPR probabilities along
with our requirement that factorizable probabilities must lead to the
classical game. This result for PD appears to diverge away from the reported
results in quantum games \cite{EWL}. We believe that part of the reason
resides with how payoff relations and players' strategies are defined in the
present framework, which exploits EPR setup for playing a quantum game.

\subsubsection{Stag Hunt}

Section \ref{SH} describes playing SH in the four-coin setup for which three
NE emerge. For each of these three NE there correspond constraints on $%
r,s,r^{\prime},s^{\prime}$ for factorizable probabilities. In the following
we first translate these constraints in terms of the EPR probabilities $%
p_{i} $ and afterwards allow $p_{i}$ to assume non-factorizable values when
the constraints on $r,s,r^{\prime},s^{\prime}$, expressed in terms of $p_{i}$%
, continue to hold. In the following we follow this procedure for each
individual NE that arises when SH is played in the four-coin setup.

\paragraph{$(x^{\star},y^{\star})_{1}=(0,0):$}

Refer to Eqs. (\ref{SH 2 coin three NE}) in Section \ref{SH} and consider
the NE $(x^{\star},y^{\star})_{1}=(0,0)$ for which the constraint on
probabilities are (\ref{s=0=s'}) as it is the case with PD. Analysis for
quantum PD from Section \ref{QPD}, therefore, remains valid and we can
directly use the inequalities (\ref{QPD Q NE 1}, \ref{QPD Q NE 2}, \ref{QPD
Q NE 3}) to obtain

\begin{equation}
\begin{array}{l}
\Pi _{A}(x^{\star },y^{\star })-\Pi _{A}(x,y^{\star })=(x^{\star }-x)\left\{
y^{\star }\triangle _{3}-\triangle _{2}(1-p_{8})/p_{1}\right\} p_{1}\geq 0,
\\ 
\Pi _{B}(x^{\star },y^{\star })-\Pi _{B}(x^{\star },y)=(y^{\star }-y)\left\{
x^{\star }\triangle _{3}-\triangle _{2}(1-p_{12})/p_{1}\right\} p_{1}\geq 0,%
\end{array}
\label{QSH (x*,y*)=(0,0)a}
\end{equation}%
where $\triangle _{3}>\triangle _{2}>0$. This gives rise to three equilibria:

\begin{equation}
\begin{array}{l}
(x^{\star },y^{\star })_{1}^{Q_{a}}=(0,0); \\ 
(x^{\star })_{2}^{Q_{a}}=(\triangle _{2}/\triangle _{3})(1-p_{12})/p_{1},%
\text{ \ \ }(y^{\star })_{2}^{Q_{a}}=(\triangle _{2}/\triangle
_{3})(1-p_{8})/p_{1}; \\ 
(x^{\star },y^{\star })_{3}^{Q_{a}}=(1,1);%
\end{array}
\label{QSH (x*,y*)=(0,0)b}
\end{equation}%
where the superscript $Q$ refers to `quantum'. From the relations (\ref{QSH
(x*,y*)=(0,0)a}, \ref{QSH (x*,y*)=(0,0)b}) and the inequalities (\ref{p1 and
p8}, \ref{p1 and p12}) it turns out that $(x^{\star },y^{\star })_{1}^{Q_{a}}
$ emerges without any further constraints apart from the ones given by
re-expressed form of (\ref{s=0=s'}) i.e. (\ref{probability value assignment}%
); $\left\{ (x^{\star })_{2}^{Q_{a}},(y^{\star })_{2}^{Q_{a}}\right\} $
emerges when non-factorizable probabilities are such that, apart from (\ref%
{probability value assignment}) to hold, both $(\triangle _{2}/\triangle
_{3})(1-p_{12})/p_{1}$ and $(\triangle _{2}/\triangle _{3})(1-p_{8})/p_{1}$
have values in the interval $[0,1]$; and $(x^{\star },y^{\star })_{3}^{Q_{a}}
$ emerges when, apart from (\ref{probability value assignment}) being true,
both $\left\{ \triangle _{3}-\triangle _{2}(1-p_{8})/p_{1}\right\} $ and $%
\left\{ \triangle _{3}-\triangle _{2}(1-p_{12})/p_{1}\right\} $ are non
negative.

\paragraph{$(x^{\star},y^{\star})_{2}=(\triangle_{2}/\triangle_{3},%
\triangle_{2}/\triangle_{3}):$}

Refer to Section \ref{SH} and use (\ref{discrete prob}) and (\ref%
{Constraints on Ps}), along with the normalization (\ref{Normalization}), to
express the constraints (\ref{SH probs constraints}) as

\begin{equation}
p_{1}=1=p_{6}\text{ and\ }p_{11}=1=p_{16}.  \label{QSH probs constraints}
\end{equation}%
The normalization (\ref{Normalization}), then, assigns zero value to the
remaining $12$ probabilities. Now substitute the constraints (\ref{QSH probs
constraints}) in Eqs. (\ref{NEx}, \ref{NEy}) to obtain the NE inequalities
that will correspond to the non-factorizable probabilities:

\begin{equation}
\begin{array}{l}
\Pi _{A}(x^{\star },y^{\star })-\Pi _{A}(x,y^{\star })=(x^{\star
}-x)[y^{\star }\triangle _{3}-\triangle _{2}]\geq 0, \\ 
\Pi _{B}(x^{\star },y^{\star })-\Pi _{B}(x^{\star },y)=(y^{\star
}-y)[x^{\star }\triangle _{3}-\triangle _{2}]\geq 0.%
\end{array}%
\end{equation}%
This results in identical to the classical situation and the three NE:

\begin{equation}
\begin{array}{l}
(x^{\star },y^{\star })_{1}^{Q_{b}}=(0,0); \\ 
(x^{\star },y^{\star })_{2}^{Q_{b}}=(\triangle _{2}/\triangle _{3},\triangle
_{2}/\triangle _{3}); \\ 
(x^{\star },y^{\star })_{3}^{Q_{b}}=(1,1);%
\end{array}
\label{QSH: (x*,y*)=delta ratio}
\end{equation}%
emerge when (\ref{QSH probs constraints}) hold.

\paragraph{$(x^{\star},y^{\star})_{3}=(1,1):$}

For this NE in (\ref{SH 2 coin three NE}) the constraint on probabilities
are (\ref{SH pure 2}) i.e. $r=0=r^{\prime}$. For factorizable probabilities,
this constraint can be rewritten using normalization (\ref{Normalization})
along with (\ref{discrete prob}, \ref{Constraints on Ps}) as

\begin{equation}
\begin{array}{l}
p_{5}=0,\text{ }p_{6}=0,\text{ }p_{9}=0,\text{ }p_{11}=0, \\ 
p_{7}+p_{8}=1=p_{10}+p_{12},\text{ }p_{4}=1,%
\end{array}
\label{SH NE3 Constraints}
\end{equation}%
from which using the normalization (\ref{Normalization}) it then follows
that $p_{1}=0,$ $p_{2}=0,$ $p_{3}=0$. The constraints (\ref{SH NE3
Constraints}) reduce the Nash inequalities (\ref{NEx}, \ref{NEy}) to

\begin{equation}
\begin{array}{l}
\Pi _{A}(x^{\star },y^{\star })-\Pi _{A}(x,y^{\star }) \\ 
=(1/2)(x^{\star }-x)[y^{\star }\triangle _{3}(2-p_{8}-p_{12}-p_{14}-p_{15})-
\\ 
\{\triangle _{3}(1-p_{8}-p_{14}-p_{15})+(\triangle _{1}+\triangle
_{2})(-1+p_{12})\}]\geq 0;%
\end{array}
\label{SH NE3 (A)}
\end{equation}

\begin{equation}
\begin{array}{l}
\Pi _{B}(x^{\star },y^{\star })-\Pi _{B}(x^{\star },y) \\ 
=(1/2)(y^{\star }-y)[x^{\star }\triangle _{3}(2-p_{8}-p_{12}-p_{14}-p_{15})-
\\ 
\{\triangle _{3}(1-p_{12}-p_{14}-p_{15})+(\triangle _{1}+\triangle
_{2})(-1+p_{8})\}]\geq 0.%
\end{array}
\label{SH NE3 (B)}
\end{equation}%
Using the constraints (\ref{SH NE3 Constraints}) in the $7$-th Equation in (%
\ref{dependent probabilities}) results in $%
p_{13}=(2-p_{8}-p_{12}-p_{14}-p_{15})/2$ which then simplifies the Nash
inequalities (\ref{SH NE3 (A)}, \ref{SH NE3 (B)}) to

\begin{equation}
\begin{array}{l}
\Pi _{A}(x^{\star },y^{\star })-\Pi _{A}(x,y^{\star })=(x^{\star }-x)\left\{
-(1-y^{\star })p_{13}\triangle _{3}+(1-p_{12})\triangle _{2}\right\} \geq 0,
\\ 
\Pi _{B}(x^{\star },y^{\star })-\Pi _{B}(x^{\star },y)=(y^{\star }-y)\left\{
-(1-x^{\star })p_{13}\triangle _{3}+(1-p_{8}-p_{13})\triangle _{2}\right\}
\geq 0,%
\end{array}
\label{SH NE r=0=r'}
\end{equation}%
and gives rise to three NE that are described below.

$\boldsymbol{(x}^{\star}\boldsymbol{,y}^{\star}\boldsymbol{)}_{1}^{Q_{c}}%
\boldsymbol{:}$ For this NE we have $(x^{\star},y^{\star})_{1}^{Q_{c}}=(0,0)$
and the inequalities (\ref{SH NE r=0=r'}) become

\begin{equation}
\begin{array}{l}
\Pi _{A}(0,0)-\Pi _{A}(x,0)=xp_{13}\triangle _{3}\left\{ 1-(\triangle
_{2}/\triangle _{3})(1-p_{12})/p_{13}\right\} \geq 0, \\ 
\Pi _{B}(0,0)-\Pi _{B}(0,y)=yp_{13}\triangle _{3}\left\{ 1-(\triangle
_{2}/\triangle _{3})(1-p_{8}-p_{13})/p_{13}\right\} \geq 0,%
\end{array}%
\end{equation}%
where from (\ref{p1 and p12}, \ref{p8 and p13}) we have $(1-p_{12})\geq 0$
and $(1-p_{8}-p_{13})\geq 0$. That is, $(x^{\star },y^{\star
})_{1}^{Q_{c}}=(0,0)$ will be a NE when $p_{8},p_{12},$ and $p_{13}$ are
such that

\begin{equation}
1\geq (\triangle _{2}/\triangle _{3})(1-p_{12})/p_{13},\text{ \ \ }1\geq
(\triangle _{2}/\triangle _{3})(1-p_{8}-p_{13})/p_{13},
\label{QSH r=0=r' (0,0)}
\end{equation}%
hold true along with the constraints (\ref{SH NE3 Constraints}).

$\boldsymbol{(x}^{\star }\boldsymbol{,y}^{\star }\boldsymbol{)}_{2}^{Q_{c}}%
\boldsymbol{:}$ From the inequalities (\ref{SH NE r=0=r'}) the strategy pair 
$(x^{\star },y^{\star })_{2}^{Q_{c}}=(x^{\star },y^{\star })$ where the
strategy pair $x^{\star }=1-(\triangle _{2}/\triangle
_{3})(1-p_{8}-p_{13})/p_{13},$ $y^{\star }=1-(\triangle _{2}/\triangle
_{3})(1-p_{12})/p_{13}$ can exist as a NE when $p_{8},p_{12},$ and $p_{13}$
are such that $x^{\star },y^{\star }\in \lbrack 0,1]$. Together with this
the constraints (\ref{SH NE3 Constraints}) are to hold.

$\boldsymbol{(x}^{\star}\boldsymbol{,y}^{\star}\boldsymbol{)}_{3}^{Q_{c}}%
\boldsymbol{:}$ For the possible NE $(x^{\star},y^{\star})_{3}^{Q_{c}}=(1,1)$
the inequalities (\ref{SH NE r=0=r'}) become

\begin{equation}
\begin{array}{l}
\Pi _{A}(x^{\star },y^{\star })-\Pi _{A}(x,y^{\star
})=(1-x)(1-p_{12})\triangle _{2}\geq 0, \\ 
\Pi _{B}(x^{\star },y^{\star })-\Pi _{B}(x^{\star
},y)=(1-y)(1-p_{8}-p_{13})\triangle _{2}\geq 0,%
\end{array}
\label{QSH r=0=r' (0,0)c}
\end{equation}%
which are to hold along with that the constraints (\ref{SH NE3 Constraints})
being true. Using $\triangle _{1},\triangle _{1}>0$ with (\ref{p1 and p12}, %
\ref{p8 and p13}) we find that the inequalities (\ref{QSH r=0=r' (0,0)c})
will always hold and the only requirement for the strategy pair $(x^{\star
},y^{\star })_{3}^{Q_{c}}=(1,1)$ to be a NE is that the constraints (\ref{SH
NE3 Constraints}) hold.

So that the list of possible NE, that can arise in the quantum SH, consists
of five members i.e. $(0,0)$, $(1,1)$, $\{\triangle _{2}(1-p_{12})/\triangle
_{3}p_{1},$ $\triangle _{2}(1-p_{8})/\triangle _{3}p_{1}\}$, $(\triangle
_{2}/\triangle _{3},$ $\triangle _{2}/\triangle _{3})$, and $[\{1-\triangle
_{2}(1-p_{8}-p_{13})/\triangle _{3}p_{13}\},$ $\{1-\triangle
_{2}(1-p_{12})/\triangle _{3}p_{13}\}]$. Which one, or more, from this list
are going to arise depends on the set of non-factorizable probabilities. For
example, we notice that there exist \cite{Cereceda} two sets of
non-factorizable probabilities that maximally violate the quantum prediction
of the Clauser-Holt-Shimony-Horne (CHSH) sum of correlations. The first set
is

\begin{equation}
\begin{array}{l}
p_{j}=(2+\sqrt{2})/8\text{ for all }p_{j}\in \mu , \\ 
p_{k}=(2-\sqrt{2})/8\text{ for all }p_{k}\in \upsilon ,%
\end{array}
\label{first set}
\end{equation}%
and the second set is

\begin{equation}
\begin{array}{l}
p_{j}=(2-\sqrt{2})/8\text{ for all }p_{j}\in \mu , \\ 
p_{k}=(2+\sqrt{2})/8\text{ for all }p_{k}\in \upsilon ,%
\end{array}
\label{second set}
\end{equation}%
where $\mu $ and $\upsilon $ are defined in (\ref{First set of probabilities}%
) and in (\ref{Second set of probabilities}), respectively. The
probabilities in these sets are non-factorizable because for both sets a
solution of the Eqs. (\ref{Pis written bilinearly}) will involve one or more
of the probabilities $r,s,r^{\prime },s^{\prime }$ being negative or greater
than one. Now for SH the requirement that factorizable probabilities are to
lead to classical game gives rise to three sets of constraints on EPR
probabilities given by (\ref{probability value assignment}), (\ref{QSH probs
constraints}), and (\ref{SH NE3 Constraints}). These sets of constraints
correspond to the NE $(x^{\star },y^{\star })_{1}=(0,0),$ $(x^{\star
},y^{\star })_{2}=(\triangle _{2}/\triangle _{3},\triangle _{2}/\triangle
_{3}),$ and $(x^{\star },y^{\star })_{3}=(1,1)$ respectively. Unfortunately,
the probabilities from either of the two sets (\ref{first set}, \ref{second
set}), which maximally violate the quantum prediction of the CHSH sum of
correlations, do not satisfy these constraints. Stated otherwise, the
probabilities from the sets (\ref{first set}, \ref{second set}) are in
conflict with the requirement that factorizable probabilities must lead to
the classical game of SH. However, other sets of non-factorizable
probabilities can be found that are consistent with this requirement and,
depending on the elements of a set, one or more out of five possible NE may
emerge. This situation can be described by saying that in SH
non-factorizability can lead to new NE but, unfortunately, either of the
sets (\ref{first set}, \ref{second set}) cannot be used for this purpose.

\subsubsection{Chicken game}

Refer to Section \ref{CG} and use Eqs. (\ref{Chicken game constraints on
r,s,r',s'}) express the constraints on $r,s,r^{\prime},s^{\prime}$ in this
setup. These constraints are imposed for the strategy pair $%
x^{\star}=\alpha/(\alpha+\beta)=y^{\star}$ is to be a NE. Use Eqs. (\ref%
{discrete prob}) and the normalization (\ref{Normalization}) to translate
these constraints in terms of the EPR probabilities $p_{i}$:

\begin{align}
\alpha (p_{2}+p_{4})& =\beta (p_{5}+p_{7}),
\label{QCG constraints on EPR probs (a)} \\
\alpha (p_{3}+p_{4})& =\beta (p_{9}+p_{10}).
\label{QCG constraints on EPR probs (b)}
\end{align}%
Addition and subtraction of (\ref{QCG constraints on EPR probs (b)}) to and
from (\ref{QCG constraints on EPR probs (a)}) gives

\begin{equation}
\begin{array}{l}
\alpha (p_{2}+p_{3}+2p_{4})=\beta (p_{5}+p_{7}+p_{9}+p_{10}), \\ 
\alpha (p_{2}-p_{3})=\beta (p_{5}+p_{7}-p_{9}-p_{10}),%
\end{array}
\label{Q Chicken 1,2}
\end{equation}%
and Eqs. (\ref{dependent probabilities}) then allow us to re-express Eqs. (%
\ref{Q Chicken 1,2}) in terms of the probabilities in set $\mu $, defined in
(\ref{Second set of probabilities}), to obtain

\begin{equation}
\begin{array}{l}
(\alpha /\beta -1)(1-p_{1}+p_{4})=p_{5}-p_{8}+p_{9}-p_{12}, \\ 
(1+\beta /\alpha )(-p_{14}+p_{15})=p_{5}-p_{8}-p_{9}+p_{12}.%
\end{array}
\label{Q Chicken a,b}
\end{equation}

Two probabilities can be eliminated from the inequalities (\ref{NEx}, \ref%
{NEy}) using Eqs. (\ref{Q Chicken a,b}). We select (arbitrarily) these to be 
$p_{12}$ and $p_{15}$ and express them in terms of other probabilities in
the set $\mu $ i.e.

\begin{equation}
\begin{array}{l}
p_{12}=p_{5}-p_{8}+p_{9}-(\alpha /\beta -1)(1-p_{1}+p_{4}), \\ 
p_{15}=p_{14}+\frac{2(p_{5}-p_{8})-(\alpha /\beta -1)(1-p_{1}+p_{4})}{%
1+\beta /\alpha }.%
\end{array}
\label{Quantum Chicken 1,2}
\end{equation}%
Notice that for the Chicken game, defined in (\ref{chicken game}), the
definition of $\triangle _{1,2}$ in Section \ref{PD(C)} gives

\begin{equation}
\triangle _{1}=\beta ,\text{ }\triangle _{2}=-\alpha .  \label{QCG deltas}
\end{equation}%
Now eliminate $p_{12}$ and $p_{15}$ from the inequalities (\ref{NEx}, \ref%
{NEy}) using Eqs. (\ref{Quantum Chicken 1,2}) and substitute from Eqs. (\ref%
{QCG deltas}). The inequalities (\ref{NEx}, \ref{NEy}) then read

\begin{equation}
\begin{array}{l}
\Pi _{A}(x^{\star },y^{\star })-\Pi _{A}(x,y^{\star }) \\ 
=\alpha (x^{\star }-x)[-y^{\star }\{(p_{1}-2p_{5}-p_{9}-p_{14}+p_{8})+ \\ 
(1-p_{1}+p_{4})\alpha /\beta +(p_{1}-p_{5}-p_{9}-p_{14})\beta /\alpha \}+ \\ 
\{(1-p_{5}-p_{14})-(p_{5}+p_{14})\beta /\alpha \}];%
\end{array}
\label{QChickenGameNE1}
\end{equation}

\begin{equation}
\begin{array}{l}
\Pi _{B}(x^{\star },y^{\star })-\Pi _{B}(x^{\star },y) \\ 
=\alpha (y^{\star }-y)[-x^{\star }\{(p_{1}-2p_{5}-p_{9}-p_{14}+p_{8})+ \\ 
(1-p_{1}+p_{4})\alpha /\beta +(p_{1}-p_{5}-p_{9}-p_{14})\beta /\alpha \}+ \\ 
\{(p_{1}-p_{4}-2p_{5}+2p_{8}-p_{9}-p_{14})+ \\ 
(1-p_{1}+p_{4})\alpha /\beta -(p_{9}+p_{14})\beta /\alpha \}].%
\end{array}
\label{QChickenGameNE2}
\end{equation}

Inequalities (\ref{QChickenGameNE1}, \ref{QChickenGameNE2}) ensure that for
factorizable probabilities the classical NE $x^{\star }=\alpha /(\alpha
+\beta )=y^{\star }$ comes out as the outcome of the game. What is the fate
of this equilibrium when probabilities are not factorizable? To address this
question we consider a special case when $\alpha =\beta $, for which $%
x^{\star }=1/2=y^{\star }$ is the classical mixed-strategy outcome of the
game. To obtain this outcome within the four-coin setup the constraints on $%
r,s,r^{\prime },s^{\prime }$ are given in Eqs. (\ref{Chicken constraint}).
The inequalities (\ref{QChickenGameNE1}, \ref{QChickenGameNE2}) reduce to

\begin{equation}
\begin{array}{l}
\Pi _{A}(x^{\star },y^{\star })-\Pi _{A}(x,y^{\star }) \\ 
=\alpha (x^{\star }-x)\{-y^{\star
}(1+p_{1}+p_{4}-3p_{5}+p_{8}-2p_{9}-2p_{14})+ \\ 
(1-2p_{5}-2p_{14})\};%
\end{array}
\label{QCGc}
\end{equation}

\begin{equation}
\begin{array}{l}
\Pi _{B}(x^{\star },y^{\star })-\Pi _{B}(x^{\star },y) \\ 
=\alpha (y^{\star }-y)\{-x^{\star
}(1+p_{1}+p_{4}-3p_{5}+p_{8}-2p_{9}-2p_{14})+ \\ 
(1-2p_{5}+2p_{8}-2p_{9}-2p_{14})\}.%
\end{array}
\label{QCGd}
\end{equation}%
Notice that the inequalities (\ref{SH NE r=0=r'}) do not allow either of the
strategy pairs $(x^{\star },y^{\star })=(1,0)$ and $(x^{\star },y^{\star
})=(0,1)$ to be NE. Like it has been the case with quantum SH, we now ask
which of these nine possible NE will emerge when probabilities become
non-factorizable. To answer this we refer to the set (\ref{first set}) of
probabilities and assign the value $(2+\sqrt{2})/8$ to each of the
probabilities $p_{1},$ $p_{4},$ $p_{5},$ $p_{8},$ $p_{9},$ $p_{14}$. Using
Eqs. (\ref{Quantum Chicken 1,2}) the assumption that $\alpha =\beta $ then
also assigns the same value, i.e. $(2+\sqrt{2})/8$, to both $p_{12}$ and $%
p_{15}$. The inequalities (\ref{QCGc}, \ref{QCGd}) are

\begin{equation}
\begin{array}{l}
\Pi _{A}(x^{\star },y^{\star })-\Pi _{A}(x,y^{\star })=(\alpha /\sqrt{2}%
)(x^{\star }-x)(y^{\star }-1)\geq 0, \\ 
\Pi _{B}(x^{\star },y^{\star })-\Pi _{B}(x^{\star },y)=(\alpha /\sqrt{2}%
)(y^{\star }-y)(x^{\star }-1)\geq 0,%
\end{array}
\label{QCGN(A,B)}
\end{equation}%
with the result that the strategy pairs $(x^{\star },y^{\star })=(1,1),(0,0)$
emerge as the new equilibria\footnote{%
Referring to (\ref{3 Equilibria in Chicken}) we recall that when $\alpha
=\beta $ there are three equilibria i.e. $(1,0),(0,1)$ and $(1/2,1/2)$ in
the classical Chicken game, at which they get rewarded by $(\alpha ,\beta
),(\beta ,\alpha )$ and $(\alpha \beta /(\alpha +\beta ),\alpha \beta
/(\alpha +\beta )),$ respectively. Here the first and the second entry
refers to Alice's and Bob's reward, respectively.}. So that, in this case,
the set (\ref{first set}) of non-factorizable probabilities indeed leads to
the new equilibria of the game. Using (\ref{coins payoff relations}) one
finds that at the equilibrium $(1,1)$ both players get $\alpha (2-\sqrt{2})/4
$ while at $(0,0)$ both players get $\alpha (2+\sqrt{2})/4$ as their payoffs.

Note that, when re-expressed in terms of the EPR probabilities $p_{i}$ using
Eqs. (\ref{discrete prob}), the constraints (\ref{Chicken constraint}) can
be written as

\begin{equation}
p_{1}+p_{2}+p_{9}+p_{10}=1=p_{1}+p_{3}+p_{5}+p_{7},  \label{QCG constraints}
\end{equation}%
which, of course, continue to hold for the set (\ref{first set}) when the
probabilities $p_{i}$ are allowed to be non-factorizable.

Similarly, referring to the second set (\ref{second set}), we assign the
value $(2-\sqrt{2})/8$ to the probabilities $p_{1},$ $p_{4},$ $p_{5},$ $%
p_{8},$ $p_{9},$ $p_{14}$ that appear in (\ref{QCGc}) and (\ref{QCGd}). The
Eqs. (\ref{Quantum Chicken 1,2}), with the assumption that $\alpha =\beta $,
then assign the value of $(2-\sqrt{2})/8$ both to $p_{12}$ and $p_{15}$ and
the inequalities (\ref{QCGc}, \ref{QCGd}) read

\begin{equation}
\begin{array}{l}
\Pi _{A}(x^{\star },y^{\star })-\Pi _{A}(x,y^{\star })=-(\alpha /\sqrt{2}%
)(x^{\star }-x)(y^{\star }-1)\geq 0, \\ 
\Pi _{B}(x^{\star },y^{\star })-\Pi _{B}(x^{\star },y)=-(\alpha /\sqrt{2}%
)(y^{\star }-y)(x^{\star }-1)\geq 0,%
\end{array}%
\end{equation}%
which are the same as the ones given in (\ref{QCGN(A,B)}), apart from extra
negative signs. This results in three strategy pairs $(x^{\star },y^{\star
})=(1,1),(1,0),(0,1)$ to come out as the equilibria. Once again, using (\ref%
{coins payoff relations}) one finds that at all of these three equilibria
both players get equally rewarded by the amount $\alpha (2+\sqrt{2})/4$.
Hence, while referring to (\ref{3 Equilibria in Chicken}), we find that in
this special case when $\alpha =\beta $ the set (\ref{second set}) of
probabilities leads to new equilibria of the game. Notice that, like it is
the case with the set (\ref{first set}) of probabilities, the constraints (%
\ref{QCG constraints}) continue to hold also for the set (\ref{second set})
when $p_{i}$ are allowed to be non-factorizable.

\section{\label{Discussion}Discussion}

In typical quantization procedure \cite{EWL,Cheon} of a two-player game, two
quantum bits (qubits) are in a quantum correlated (entangled) state that are
given to the players Alice and Bob. Players' strategies consist of
performing unitary actions on their respective qubits. Classical game
remains a subset of the quantum game in that both players can play quantum
strategies that correspond to the strategies available classically.

As more choices are allowed to the players that now also include
superpositions of their classical moves, it gives ground to the argument%
\footnote{%
Enk and Pike \cite{EnkPike} argument can be described as follows. The
extended set of players' moves allows us to construct an extended payoff
matrix that includes extra available moves. Enk and Pike interpret this by
saying that the `essence' of a quantized game can be captured by a different
classical game and it is the new game that is constructed and solved and 
\emph{not} the original classical game.} that Enk and Pike \cite{EnkPike}
have put forward. The setup proposed in this paper uses EPR experiments to
play a two-player game and a quantum game is associated to a classical game
such that it becomes hard to construct an Enk and Pike-type argument as both
the payoff relations and the players' sets of strategies remain identical 
\cite{IqbalWeigert} in the classical and the associated quantum game.

In the present setup it is non-factorizability -- responsible for the
violation of Bell inequality in EPR experiments -- that gives rise to the
new solutions in quantum game. When players play a game using a physical
system for which joint probabilities are factorizable the classical game
results always. In other words, the role of non-factorizable probabilities
is sought in the game-theoretic solution concept of a NE, when the physical
realization for these probabilities is provided by the EPR experiments. This
analysis introduces a new viewpoint in the area of quantum games in which
non-factorizability gets translated into the language of game theory.

The argument put forward in this paper can be described as follows. Firstly,
players' payoffs are re-expressed in the form $\Pi_{A,B}(p_{i},x,y,\mathcal{A%
},\mathcal{B})$ where $p_{i}$ are the sixteen joint probabilities; $x,y$ are
players strategies, and $\mathcal{A},\mathcal{B}$ are players' payoff
matrices defined in (\ref{A and B matrices}). Secondly, Nash inequalities
are used to impose constraints on $p_{i}$ that ensure that with factorizable 
$p_{i}$ the game has classical outcome and the resulting payoffs can be
interpreted in terms of classical mixed-strategy game. It is achieved by
playing the game in the four-coin setup and using Nash inequalities to
obtain constraints on the coin probabilities $r,s,r^{\prime},s^{\prime}$
which reproduce the outcome of the classical mixed-strategy game. Using (\ref%
{discrete prob}), which results from factorizability, these constraints on $%
r,s,r^{\prime},s^{\prime}$ are then translated in terms of constraints on $%
p_{i}$. Thirdly, while referring to the EPR setup, $p_{i}$ are allowed to be
non-factorizable when the constraints on $p_{i}$ continue to hold. Fourthly,
and lastly, it is observed how non-factorizability leads to the emergence of
new solutions of the game.

Note that for a game different sets of constraints are defined depending on
which NE is to be the solution of the game. For example, for three NE in
Chicken we require three different sets of constraints on $r,s,r^{\prime
},s^{\prime}$. Considering one of these three sets at a time we repeat the
four steps stated above. The same procedure is then repeated for other sets
of constraints corresponding to other NE.

That is, in this setup not all solutions of a game are re-expressed in terms
of a single set of constraints on $r,s,r^{\prime},s^{\prime}$. Instead, a
separate set of constraints is found for each NE. It seems that the
four-coin setup is the minimal arrangement that allows one to introduce, in
a smooth way, the EPR probabilities into a game-like setting. We suggest
that with increasing the number of coins, shared by each of the two players,
all the NE of a game can be translated to a single constraint on the
underlying coin probabilities which are subsequently translated in terms of $%
p_{i}$. This will then allow to see the role of non-factorizability on
solution of a game from a single set of constraints. However, this will be
obtained at a price: Firstly, more coins will be involved resulting in more
mathematical complexity; secondly, for more coins player's strategy will
need to be redefined such that it permits to incorporate EPR probabilities.

Note that the usual approach uses entangled states to construct quantum
games and this paper uses non-factorizability to the same end.
Mathematically, non-factorizability comes out to be a stronger condition
than the condition that translates entanglement into constraints on joint
probabilities. That is, a non-factorizable set of probabilities always
corresponds to some entangled state but an entangled state can produce a
factorizable set of probabilities. For example, in case of singlet state the
outcomes of two measurements violate Bell inequality only along certain
directions, and not along other directions. In other words, in a quantum
game exploiting entangled states, the joint probabilities may still be
factorizable but for a quantum game, resulting from non-factorizable
probabilities, Bell inequality is bound to be violated. Non-factorizability
being a stronger condition may well be suggested as the reason why it cannot
be helpful to escape from the classical outcome in PD.

The proposed setup demonstrates how non-factorizability can change outcome
of a game. We suggest to extend \cite{IqbalCheon} this setup to analyze
multi-player quantum games \cite{Multiplayer} where players share physical
systems for which joint probabilities cannot be factorized.

\begin{acknowledgement}
Authors are thankful to anonymous referees for their helpful comments.
\end{acknowledgement}

\begin{acknowledgement}
One of us (AI) is supported by the Japan Society for Promotion of Science
(JSPS). AI gratefully acknowledges hospitality and support from the
International Relations Center and the Research Institute at the Kochi
University of Technology. This work is supported, in part, by the
Grant-in-Aid for scientific research provided by Japan Ministry of
Education, Culture, Sports, Science, and Technology under the contract
numbers 18540384 and 18.06330.
\end{acknowledgement}


\begin{thebibliography}{99}
\bibitem{MeyerDavid} D. A. Meyer, Quantum Strategies, Phys. Rev. Lett. 
\textbf{82,} 1052 (1999).

\bibitem{EWL} J. Eisert, M. Wilkens, and M. Lewenstein, Quantum Games and
Quantum Strategies. Phy. Rev. Lett. \textbf{83}, 3077 (1999). Also, J.
Eisert and M. Wilkens, Quantum Games. J. Mod. Opt. \textbf{47}, 2543 (2000).

\bibitem{Multiplayer} S. C. Benjamin, P. M. Hayden, Multi-Player Quantum
Games, Phys. Rev. A \textbf{64}, 030301 (2001).

\bibitem{Du} J. Du, Hui Li, X. Xu, M. Shi, J. Wu, X. Zhou and R. Han,
Experimental Realization of Quantum Games on a Quantum Computer, Phys. Rev.
Lett. \textbf{88}, 137902 (2002).

\bibitem{Cheon} T. Cheon, I. Tsutsui, Classical and Quantum Contents of
Solvable Game Theory on Hilbert Space, Phys. Lett. A\textbf{348}, 147 (2006).

\bibitem{Flitney} A. Flitney, Aspects of Quantum Game Theory, PhD thesis,
University of Adelaide (2005), and the references therein.

\bibitem{Shimamura} J. Shimamura, S. K. \"{O}zdemir, F. Morikoshi and N.
Imoto, Quantum and Classical Correlations Between Players in Game Theory,
Int. J. Quant. Inf. \textbf{2/}1, 79 (2004).

\bibitem{Iqbal} A. Iqbal, Studies in the Theory of Quantum Games, PhD
thesis, Quaid-i-Azam University (2004), and the references therein. \texttt{%
quant-ph/0503176}.

\bibitem{Ichikawa} T. Ichikawa, I. Tsutsui, and T. Cheon, Quantum Game
Theory Based on the Schmidt Decomposition: Can Entanglement Resolve
Dilemmas?, \texttt{quant-ph/0702167}

\bibitem{Peres} A. Peres, \textit{Quantum Theory: Concepts and Methods},
Kluwer Academic Publishers (1995).

\bibitem{Nielsen} M. A. Nielsen and I. L. Chuang, \textit{Quantum
Computation and Quantum Information}, Cambridge University Press (2000). See
also, C. P. Williams and S. H. Clearwater, \textit{Explorations in Quantum
Computing}, Springer-Verlag Inc, New York (1998).

\bibitem{Rasmusen} E. Rasmusen, \textit{Games \& Information}: \textit{An
Introduction to Game Theory}, 3rd Edition, Blackwell publishers Ltd, Oxford,
(2001).

\bibitem{Chung} See for example K. L. Chung and F. AitSahlia, \textit{%
Elementary Probability Theory}, Fourth Edition, Springer-Verlag New York
(2003).

\bibitem{Bell} J. S. Bell, On the Einstein-Poldolsky-Rosen Paradox, Physics 
\textbf{1}, 195 (1964). Also by the same author, On the Problem of Hidden
Variables in Quantum Mechanics, Rev. Mod. Phys. \textbf{38}, 447452 (1966).

\bibitem{Bell1} J. S. Bell, \textit{Speakable and Unspeakable in Quantum
Mechanics}, Cambridge University Press, Cambridge, (1987).

\bibitem{Bell2} J. F. Clauser and A. Shimony, Bells Theorem: Experimental
Tests and Implications, Rep. Prog. Phys. \textbf{41}, 1881 (1978).

\bibitem{EPR} A. Einstein, B. Podolsky and N. Rosen, Can Quantum-Mechanical
Description of Physical Reality Be Considered Complete?, Phys. Rev. \textbf{%
47}, 777 (1935).

\bibitem{Aspect et al} A. Aspect, J. Dalibard, and G. Roger, Experimental
Test of Bell's Inequalities Using Time-Varying Analyzers, Phys. Rev. Lett. 
\textbf{49}, 1804 (1982).

\bibitem{Aumann} R. J. Aumann and S. Hart (editors), \textit{Handbook of
Game Theory with Economic Applications}, Volume 1, published by Elsevier
(1992).

\bibitem{Hofbauer Sigmund} J. Hofbauer and K. Sigmund, \textit{Evolutionary
Games and Population Dynamics}, Cambridge University Press, Cambridge (1988).

\bibitem{Fine} A. Fine, \textit{The Shaky Game: Einstein, Realism, and the
Quantum Theory}, Second Edition, The University of Chicago Press, (1996),
page 59.

\bibitem{WinsbergFine} E. Winsberg and A. Fine, Quantum Life: Interaction,
Entanglement, and Separation, Journal of Philosophy C \textbf{80} (2003).

\bibitem{Fine1} See Ref. \cite{Fine} page 63.

\bibitem{Iqbal(1)} A. Iqbal, Playing Games with EPR-Type Experiments, J.
Phys. A: Math. Gen. \textbf{38,} 9551 (2005).

\bibitem{Clauser Horne} J. F. Clauser and M. A. Horne, Experimental
Consequences of Objective Local Theories, Phy. Rev. D, Vol. \textbf{10},
Number 2, 526 (1974).

\bibitem{Cereceda} J. L. Cereceda, Quantum Mechanical Probabilities and
General Probabilistic Constraints for Einstein-Podolsky-Bohm Experiments,
Found. Phys. Lett. Vol. \textbf{13}, No. 5, (2000).

\bibitem{EnkPike} S. J. van Enk and R. Pike, Classical Rules in Quantum
Games, Phys. Rev. A \textbf{66}, 024306 (2002).

\bibitem{IqbalWeigert} A. Iqbal and S. Weigert, Quantum Correlation Games,
J. Phys. A, Math. Gen. \textbf{37,} 5873 (2004).

\bibitem{IqbalCheon} A. Iqbal and T. Cheon, Constructing Multi-Player
Quantum Games from Non-Factorizable Joint Probabilities, in writing
\end{thebibliography}
\end{document}